\documentclass[fleqn,usenatbib]{mnras}

\usepackage{newtxtext,newtxmath}
\usepackage{bm}

\usepackage{tablefootnote}

\usepackage[T1]{fontenc}

\DeclareRobustCommand{\VAN}[3]{#2}
\let\VANthebibliography\thebibliography
\def\thebibliography{\DeclareRobustCommand{\VAN}[3]{##3}\VANthebibliography}

\usepackage{graphicx}	
\usepackage{amsmath}	

\newcommand{\psr}{1E~1207.4$-$5209}
\newcommand{\XMM}{\textit{XMM-Newton}}
\newcommand{\Chandra}{\textit{Chandra}}
\newcommand{\NICER}{\textit{NICER}}
\newcommand{\NH}{N_{\rm H}}

\newcommand{\sigmami}{\sigma_{n-}}
\newcommand{\sigmapi}{\sigma_{n+}}
\newcommand{\linedi}{\delta_n}
\newcommand{\enone}{E_1}
\newcommand{\entwo}{E_2}
\newcommand{\enthree}{E_3}
\newcommand{\sigmamone}{\sigma_{1-}}
\newcommand{\sigmamtwo}{\sigma_{2-}}
\newcommand{\sigmamthree}{\sigma_{3-}}
\newcommand{\sigmapone}{\sigma_{1+}}
\newcommand{\sigmaptwo}{\sigma_{2+}}
\newcommand{\sigmapthree}{\sigma_{3+}}
\newcommand{\linedone}{\delta_1}
\newcommand{\linedtwo}{\delta_2}
\newcommand{\linedthree}{\delta_3}
\newcommand{\Eb}{E_B}

\title[Two decades of X-ray spectra of \psr]{Searching for magnetic field changes in the central compact object \psr\ in two decades of X-ray spectra }

\author[T. Leon and W. C. G. Ho]{
Timothy Leon,$^{1}$
and
Wynn C. G. Ho,$^{1}$
\\
$^{1}$Department of Physics and Astronomy, Haverford College, 370 Lancaster Avenue, Haverford, PA 19041, USA
}

\date{Accepted 2026 November 10. Received 2026 November 9; in original form 2026 August 19}

\pubyear{2026}

\begin{document}
\label{firstpage}
\pagerange{\pageref{firstpage}--\pageref{lastpage}}
\maketitle

\begin{abstract}
The central compact object (CCO) \psr\ is remarkable in that its X-ray spectrum
shows clear absorption features at $\approx0.7$ and 1.4~keV.
Attributing these features to electron cyclotron resonance, the implied surface
magnetic field is $B\approx(7-8)\times10^{10}\mbox{ G}$, which agrees
approximately with the field inferred from the spin properties of the pulsar.
Here we examine all \XMM\ EPIC~pn spectra from
more than 30 observations of \psr\ from 2001--2020.
We model the rotation phase-averaged and phase-resolved spectra with a two
blackbody and multiple absorption line model, focusing on
potential changes in the absorption lines which would correspond
to changes in the magnetic field that could be related to the
pulsar's possible spin glitch in 2015 and recent detection in radio.
We find the phase-averaged spectral parameters, and by implication the magnetic field, are constant across nearly 20~years of data, while some parameters characterizing the phase-resolved spectra vary with rotation phase.
The spectral line energies and widths (e.g., $\enone\approx0.66\mbox{ keV}$ and full width of $0.34\mbox{ keV}$) imply the magnetic field distribution across the X-ray emitting regions spans the range $B\approx(6-11)\times10^{10}\mbox{ G}$, with the highest fields being visible during the declining phase of the pulsations of \psr.
\end{abstract}


\section{Introduction} \label{sec:intro}

Approximately a dozen neutron stars (NS) are currently classified as central compact objects (CCOs). These CCOs are located near the centre of a supernova remnant and mainly emit only soft thermal X-ray emission \citep{Deluca2008, Deluca2017, Gotthelf2013}. Only 3 of the known CCOs have pulsations detected from them: PSR~J0821$-$4300, PSR~J1210$-$5226 (also known as \psr), and PSR~J1852+0040. The spin period of each  CCO is $P = (0.112,0.424,0.105)\mbox{ s}$, respectively
\citep{Zavlin2000,Gotthelf2005,Gotthelf2009},
and their low spin-down rate is
$\dot{P} = (7.0, 22, 8.7) \times 10^{-18}\mbox{ s s$^{-1}$}$, respectively
\citep{Halpern2010,Gotthelf2013}.
The surface dipole magnetic field strengths of the pulsars can be estimated using $B = 3.2 \times 10^{19} \sqrt{P\dot{P}}$, resulting in $B = (2.9,9.8,3.1) \times 10^{10}\mbox{ G}$, respectively. These fields are one to two orders of magnitude lower than what is typical for young NSs.

\psr\ is a CCO in the supernova remnant PKS~J1209$-$51/52 (also known as G296.5+10; \citealt{Helfand1984,Kellett1987,Mereghetti1996,Vasisht1997}) and is at a distance and age of about 1.4~kpc and 14~kyr, respectively \citep{Eppens2024}.  It is particularly unusual due to its X-ray spectrum having clear absorption lines at 0.7 and 1.4~keV \citep{Mereghetti2002,Sanwal2002,Bignami2003,Mori2005}. If one interprets these lines as being due to electron cyclotron harmonics, with the fundamental occurring at the redshifted energy
\begin{equation}
\Eb = 0.116\mbox{ keV}(B/10^{10}\mbox{ G})/(1+z_g), \label{eq:cyc}
\end{equation}
where $1+z_g$ is the gravitational redshift factor, then a magnetic field $B = (7-8) \times 10^{10}\mbox{ G}$ is implied, which is similar to the $B = 9.8 \times 10^{10}\mbox{ G}$ inferred from its spin period and period derivative. This independent measurement of magnetic field of \psr, as well as those possibly in PSR~J0821$-$4300 \citep{Gotthelf2009,Gotthelf2013} and the Vela~Junior CCO \citep{Ho2026}, confirms the abnormally low surface magnetic field strength of CCOs.

\psr\ also appeared to undergo a glitch in 2015 \citep{Gotthelf2018,Gotthelf2020}.
Glitches are brief periods of increased (or decreased in a few instances) rotation due to the unpinning of superfluid
vortices \citep{Anderson1975,Alpar1984,Antonopoulou2022} and are correlated with high $\dot{P}$ \citep{Espinoza2011,Fuentes2017}.  Thus the glitch of \psr\ was quite unexpected
but continues to be supported by a longer post-glitch baseline of data, which includes measurement of an unusually large and negative $\Delta\dot{f}/\dot{f}=-0.09$, although extreme timing noise could provide an alternative explanation (see \citealt{Perez2026}, and discussion therein).
A negative change in spin-down rate of 9~percent at the epoch of the glitch would suggest a decrease in dipolar magnetic field strength of about 4.5~percent, or $\sim4.5\times10^9\mbox{ G}$.
On the other hand, \citet{Gotthelf2018} analysed phase-averaged spectra from 2012 and 2017,
i.e., before and after the glitch, and found no evidence for any spectral changes
and hence no indication of changes in the magnetic field of \psr.

Further evidence for transient behavior in \psr\ is its recent observation in radio
\citep{Zhang2026}, in contrast to earlier searches and non-detections
(see, e.g., \citealt{Kaspi1996,Cotton2024,Turner2024}).
This represents the first radio detection of a CCO and challenges the long-standing picture of CCOs as radio-quiet objects. The emergence of  radio activity may indicate changes in the magnetospheric environment of \psr, spurred by possible changes in its surface magnetic field.

In this paper, we present spectral analyses of \XMM\ observations of \psr.  Our objective is to search for temporal variations in the cyclotron absorption features.
We confine our attention to \XMM\ data and neglect \Chandra\ and \NICER\ data.
This allows us to avoid additional uncertainties due to calibration
differences between these instruments and varying spectral contamination from
the supernova remnant since \Chandra\ observations were taken in continuous
clocking mode and hence only have spatial resolution in one dimension and
\NICER\ is non-imaging.
Previous works examining \XMM\ spectra studied the 2001 \citep{Mereghetti2002}, 2002 \citep{Bignami2003,DeLuca2004,Mori2005}, 2012 and 2017 \citep{Gotthelf2018}, or 2017 dataset \citep{Alford2023}.  Here we examine all observations from 2001 to 2020.  In Section~\ref{sec:dataanalysis}, we describe these \XMM\ data and our assumed spectral model. In Section~\ref{sec:results}, we discuss the results of our phase-averaged and phase-resolved analyses. We discuss the implications of our findings in Section~\ref{sec:discuss} and conclude in Section~\ref{sec:conclusion}.

\begin{table}
\centering
\caption{Log of all \XMM\ observations of \psr.}
\label{tab:xmm_obs}
\setlength{\tabcolsep}{5pt}
\begin{tabular}{ccccc}
\hline
Dataset & Date & ObsID & Exposure & Counts \\
Name &&& (ks) & ($10^3$) \\
\hline
& 2001 Dec 23 & 0113051001$^\ast$ & 1 \\
2001 & 2001 Dec 23 & 0113050501 & 19 & 24 \\
2002&&&152 & 201 \\
 & 2002 Aug 4 & 0155960301 & 77 \\
& 2002 Aug 6 &0155960501 & 75 \\
2005 & & & 84 & 107 \\
& 2005 Jun 22 & 0304531501 & 10 \\
     & 2005 Jul 5 & 0304531601  & 13 \\
     & 2005 Jul 10 & 0304531701  & 5 \\
     & 2005 Jul 11 & 0304531801  & 40 \\
     & 2005 Jul 12 & 0304531901  & 3 \\
     & 2005 Jul 17 & 0304532001  & 10 \\
     & 2005 Jul 31 & 0304532101  & 2 \\
2008 & & & 43 & 56 \\
& 2008 Jul 2 & 0552810301 & 22\\
     & 2008 Dec 22 & 0552810401 & 21 \\
2012 & & & 102 & 118 \\
& 2012 Jun 22 & 0679590101 & 19\\
     & 2012 Jun 24 & 0679590201 & 16 \\
     & 2012 Jun 28 & 0679590301 & 17 \\
     & 2012 Jul 2 & 0679590401 & 12 \\
     & 2012 July 18 & 0679590501 & 19 \\              
     & 2012 Aug 11 & 0679590601 & 19 \\
2016 & 2016 Jul 28 & 0780000201 & 23 & 30 \\
2017 & & & 80 & 105 \\
& 2017 Jun 22 & 0800960201 & 19\\
     & 2017 Jun 23 & 0800960301 & 15 \\
     & 2017 Jun 24 & 0800960401 & 16 \\
     & 2017 Jul 3 & 0800960501 & 16 \\
     & 2017 Aug 10 & 0800960601 & 14 \\
2018 & & & 66 & 85 \\
& 2017 Dec 24 & 0800960701 & 13 \\
     & 2018 Jun 22 & 0821940201 & 22\\
     & 2018 Dec 28 & 0821940301 & 15\\
     & 2019 Jul 9 & 0842280301 & 16\\
2020 & & & 54 & 70 \\
& 2020 Jan 16 & 0842280601 & 21\\
     & 2020 Jun 22 & 0862990201 & 23 \\
     & 2020 Dec 25 & 0862990301 & 10\\
\hline
\multicolumn{5}{l}{$^\ast$ Not used.}
\end{tabular}
\end{table}

\section{Data Analysis} \label{sec:dataanalysis}
\subsection{\XMM\ Data} \label{sec:data}
\subsubsection{Phase-Averaged Spectroscopy} \label{sec:dataavg}
Over the course of nearly 20~years, \XMM\ EPIC observed \psr\ on 32~separate occasions.  A log of these is given in Table~\ref{tab:xmm_obs}.  We exclude from our analysis ObsID~0113051001 from 2001, due to its short effective exposure time.

For this work, we use only data from EPIC pn \citep{Struder2001} and not MOS \citep{Turner2001} since pn has better spectral sensitivity and a time resolution of 5.7~ms capable of measuring the pulsations of \psr\ while MOS does not.  We process pn observation data files (ODFs) with the Science Analysis Software (SAS)~22.0.0 task \texttt{epproc}. To remove background flaring in the observations, we extract single-event (PATTERN~$=0$), high-energy (10--12 keV) light curves and determine a steady count rate threshold to be between 0.1--0.15~$\mathrm{c}\,\mathrm{s^{-1}}$,  depending on the observation. We generate good time intervals (GTIs) via \texttt{tabgtigen} and produce flare-cleaned event files with \texttt{evselect}. This results in effective exposure times for the datasets shown in Table~\ref{tab:xmm_obs}.

To extract source and background counts, we use circular regions of radius 30~arcsec and 50~arcsec, respectively. The background region is chosen to be an area with no obvious sources and uniformly low counts. We use \texttt{epatplot} and do not find our source counts to be significantly affected by pileup. For spectral analysis, source and background counts are extracted using FLAG~$=0$ and PATTERN~$\leq 4$. Effective areas are calculated using \texttt{backscale}. To compute RMF and ARF files, we use \texttt{rmfgen} and \texttt{arfgen}, respectively.

In order to improve
the statistics of our spectral modelling fits, we use \texttt{epicspeccombine} to merge observations, with combinations depending on the purpose. For example, we combine all observations within 6 months of each other, unless a different grouping produces better fit statistics. The data from 2017~December (ObsID~0800960701) is kept separate from the rest of the data taken in 2017 due to poor fit results when combined. The 2016 observation (ObsID~0780000201) is also kept separate for the same reason. The resulting combined datasets at 9 epochs are indicated in Table~\ref{tab:xmm_obs}. Spectra are binned using \texttt{specgroup} to a minimum of 100~counts per bin. 

\subsubsection{Phase-Resolved Spectroscopy} \label{sec:dataphase}
To obtain phase-resolved spectra, we follow the same procedure as described above, except we restrict the energy range to 0.3--3.0~keV and use \texttt{barycen} to barycenter the events from each observation. To assign phase values to each event, we use \texttt{phasecalc} with the preglitch and postglitch ephemerides shown in Table 1 of \citet{Perez2026}.

\begin{figure}
\centering
\includegraphics[width=\columnwidth, trim=10 25 10 20, clip]{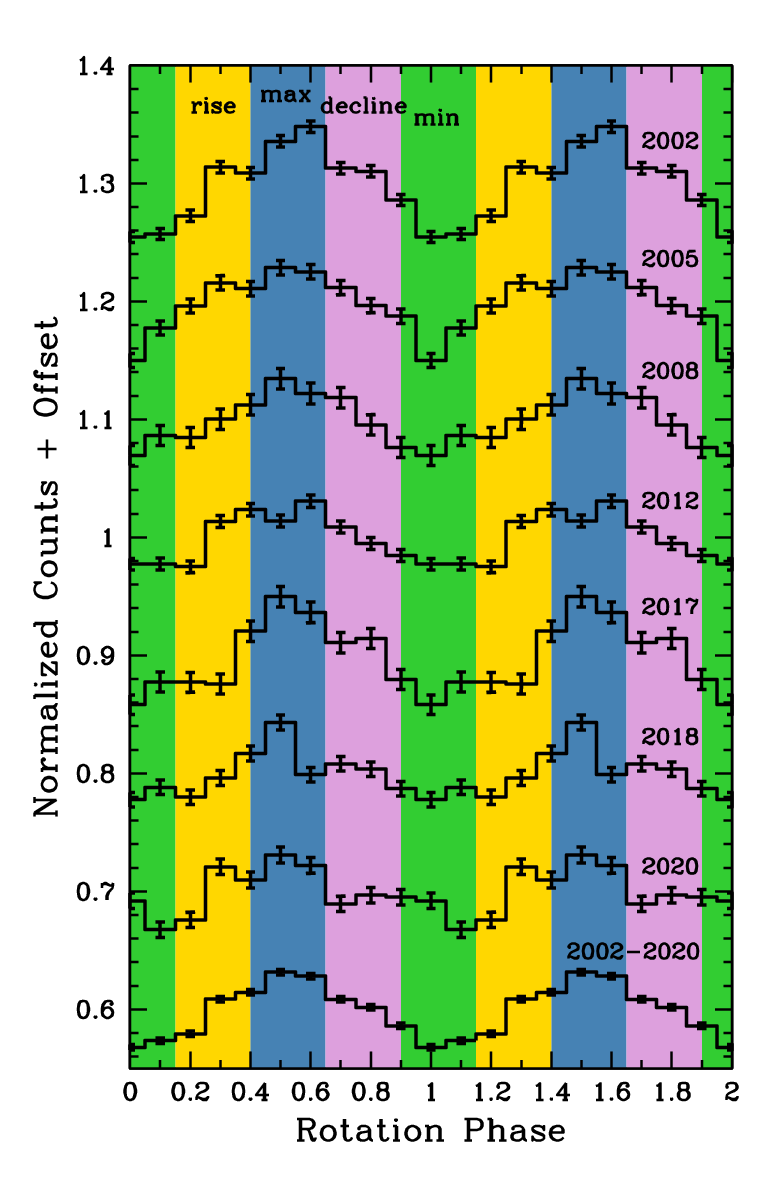}
\caption{0.3--3~keV pulse profile of \psr\ from \XMM\ EPIC pn data from 2002 to 2020 and the summed total of all epochs (top to bottom).  Each pulse profile is normalized by its average counts per bin, and arbitrary offsets are used for illustration.  Errors are 1$\sigma$ uncertainty.   Two rotation cycles are shown for clarity.  Shading indicates phase ranges for phase-resolved spectral analyses: 0.15--0.4 for rising, 0.4--0.65 for maximum, 0.65--0.9 for declining, and 0.9--1.15 for minimum. \label{fig:pp}}
\end{figure}

Since 1E 1207.4$-$5209 has a broad pulse profile, we divide it into four equal segments: maximum, declining, minimum, and rising with pulse ranges of 0.4--0.65 for maximum, 0.65--0.9 for declining, 0.9--1.15 for minimum, and 0.15--0.4 for rising. The 2001 and 2016 datasets are excluded due to their relatively low exposure times when divided into 4 segments, leaving 7 epochs of data for phase-resolved analysis. The pulse profile for each dataset is shown in Figure~\ref{fig:pp}.  The phase-resolved spectra are then grouped to a minimum of 25~counts per bin using \texttt{specgroup}. 

\subsection{Spectral Modelling} \label{sec:model}
Spectral fitting is performed using Xspec~12.15.0 \citep{Arnaud1996} with an energy range of 0.3--4 keV for phase-averaged data and 0.3--3~keV for phase-resolved data. Due to the complicated spectrum of \psr, the spectral model we use is composed of several components. To account for photoelectric absorption in the interstellar medium, we use \texttt{tbabs}, which is parametrized by the effective hydrogen column density $\NH$, with abundances from \citet{Wilm2000} and cross-sections from \citet{Verner1996}. We note that, due to difficulties in constraining the normalization for the phase-resolved data, $\NH$ for these fits are frozen to the best-fit value from the corresponding phase-averaged fits.

Similar to \citet{DeLuca2004}, we model the intrinsic spectrum of 1E 1207.4$-$5209 as a double blackbody (BB+BB) alongside absorption features at $\approx0.7\mbox{ keV}$, 1.4~keV, and, if statistically significant, 2.1~keV. Since our objective is searching for variations in the absorption features, we do not consider more physical or accurate models for the continuum spectra.  For the blackbody components, we use \texttt{bbodyrad}, which is parameterized by the temperature $kT$ and normalization~($=R^2/D^2$, where $R$ is the radius of the emitting region and $D$ is the distance).  For each absorption feature, we initially use the built-in Xspec function \texttt{gabs}, which assumes a symmetric Gaussian shape. However, previous works found indications of asymmetric shapes and improved fit results when an asymmetric shape is used (e.g., \citealt{Sanwal2002,DeLuca2004}).  Indeed, we find a symmetric Gaussian shape generally produces poor fit statistics for most datasets.
To address this, we instead define the following model to allow for an asymmetric shape (i.e., when $\sigmami\neq\sigmapi$),
\begin{equation}
    F(E)=
    \begin{cases}
    1-\linedi\cdot\exp\left[-\frac{(E-E_n)^2}{2 \sigmami^2}\right] & \text{for } E \le E_n \\
    1-\linedi\cdot\exp\left[-\frac{(E-E_n)^2}{2\sigmapi^2}\right] & \text{for } E > E_n 
    \end{cases}, \label{eq:asym}
\end{equation}
where $n=1,2,3$ for up to three absorption features centred at $E_n$, depth of the feature is $\linedi$, and widths below and above $E_n$ are $\sigmami$ and $\sigmapi$, respectively.  Note that \citet{DeLuca2004} use a different analytic form from equation~(\ref{eq:asym}) to model the asymmetric absorption features.
When relating spectral line energies to the magnetic field $B$ via equation~(\ref{eq:cyc}), we assume for simplicity a gravitational redshift factor $1+z_g=1.25$;
we note, as an example, that a NS with a mass of $1.4\,M_\odot$ and radius of 11.5~km has a redshift factor of 1.25.

An asymmetric "Gaussian" line, like that given by equation~(\ref{eq:asym}), might be generated by the following.
Consider a circular spot on the NS surface, e.g., one of the magnetic poles, and assume for simplicity it is uniform in temperature.
The magnetic field strength across this spot has a small range of values, from high in the center to lower on the edge, and this range of $B$ would produce cyclotron absorption lines with a small range in energy, according to equation~(\ref{eq:cyc}).
A symmetric Gaussian-shaped spectral line might then be produced if the largest portion of the circular spot (i.e., annulus with the largest area) has the range's median magnetic field, and thus the line is centered on the energy that corresponds to this median $B$ while the width of the line is determined by the range of $B$.
An asymmetric Gaussian would be the result of the largest area not having the median $B$.
Note that the intrinsic shape of a single cyclotron line is Lorentzian,
higher cyclotron harmonics generally manifest as weaker, narrower spectral
lines than the fundamental line,
and the above is a very simplified scenario, where we ignore other line
broadening effects (see, e.g.,
\citealt{Pavlov1978,Potekhin2010,Suleimanov2010,Suleimanov2012,Potekhin2014}).
A more detailed study will be conducted in future work.

\begin{figure}
\centering
\includegraphics[width=\columnwidth, trim=0 20 15 15, clip]{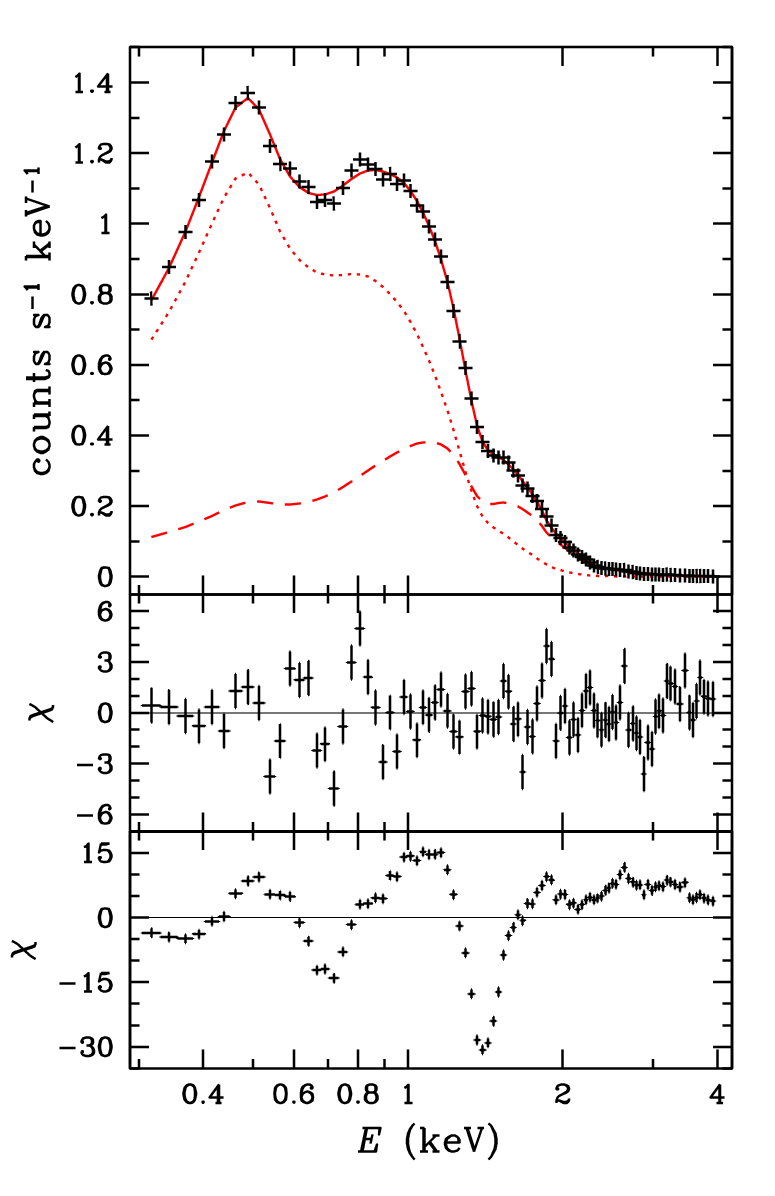}
	\caption{\XMM\ EPIC pn spectrum of \psr.  Top panel shows data combined from all observations from 2001--2020 with $1\sigma$ errors (crosses).  Solid line shows spectral model made up of two blackbodies and three spectral lines, with best-fitting parameters given in Table~\ref{tab:avgall}, and the dashed and dotted lines show the small hot and large cool blackbody components, respectively.
Middle panel shows relative residual, $\chi=(\mbox{data}-\mbox{model})/\mbox{error}$, from the top panel results.
For comparison, the bottom panel shows $\chi=(\mbox{data}-\mbox{model})/\mbox{error}$ of a best-fit two blackbody model \textit{without} spectral lines.
\label{fig:avgall}}
\end{figure}

\begin{table*}
\renewcommand{\arraystretch}{1.3}
    \centering
    \caption{Phase-averaged and phase-resolved (Max, Decline, Min, and Rise phase ranges, as shown in Figure~\ref{fig:pp}) spectral modelling results, where the model is composed of two blackbodies and two or three absorption lines and fits are to combined spectra from all observations.
    $R_1$ and $R_2$ are calculated from the blackbody normalizations assuming a distance $D=1.4\mbox{ kpc}$.
    The absorption line model has a Gaussian shape when line widths are equal, i.e., $\sigmapi=\sigmami$.
    Errors are $1\sigma$.}
    \label{tab:avgall}
    \begin{tabular}{|lccccr} 
        \hline
        Parameter& Phase-averaged &  Max & Decline & Min & Rise \\ \hline
$\NH$ ($10^{21}\mbox{ cm$^{-2}$}$) & $0.88_{-0.04}^{+0.05}$ & $0.97_{-0.10}^{+0.11}$ & $1.19_{-0.21}^{+0.16}$ & $0.75_{-0.08}^{+0.09}$ & $1.11_{-0.09}^{+0.10}$ \\
$kT_1$ (keV) & $0.184_{-0.005}^{+0.005}$ & $0.178_{-0.012}^{+0.018}$ & $0.194_{-0.012}^{+0.016}$ & $0.184_{-0.008}^{+0.008}$ & $0.159_{-0.005}^{+0.006}$ \\
$R_1$ (km) & $2.13_{-0.65}^{+0.75}$ & $2.2_{-1.1}^{+1.4}$ & $2.9_{-1.9}^{+2.6}$ & $2.01_{-0.80}^{+0.95}$ & $2.9_{-1.2}^{+1.4}$ \\
$kT_2$ (keV) & $0.328_{-0.005}^{+0.005}$ & $0.310_{-0.011}^{+0.029}$ & $0.380_{-0.043}^{+0.065}$ & $0.328_{-0.010}^{+0.012}$ & $0.316_{-0.006}^{+0.007}$ \\
$R_2$ (km) & $0.44_{-0.16}^{+0.17}$ & $0.54_{-0.39}^{+0.32}$ & $0.22_{-0.18}^{+0.28}$ & $0.43_{-0.22}^{+0.22}$ & $0.53_{-0.20}^{+0.21}$ \\
$\enone$ (keV) & $0.662_{-0.004}^{+0.004}$ & $0.702_{-0.004}^{+0.009}$ & $0.689_{-0.009}^{+0.006}$ & $0.637_{-0.008}^{+0.007}$ & $0.677_{-0.006}^{+0.007}$ \\
$\sigmamone$ (keV) & $0.083_{-0.006}^{+0.007}$ & $0.12_{-0.02}^{+0.02}$ & $0.19_{-0.06}^{+0.07}$ & $0.06_{-0.01}^{+0.01}$ & $0.08_{-0.01}^{+0.01}$ \\
$\sigmapone$ (keV) & $0.25_{-0.01}^{+0.01}$ & $0.21_{-0.02}^{+0.04}$ & $0.46_{-0.07}^{+0.08}$ & $0.25_{-0.01}^{+0.01}$ & $0.168_{-0.009}^{+0.010}$ \\
$\linedone$ (keV) & $0.448_{-0.008}^{+0.009}$ & $0.39_{-0.02}^{+0.03}$ & $0.65_{-0.10}^{+0.09}$ & $0.51_{-0.01}^{+0.01}$ & $0.47_{-0.01}^{+0.01}$ \\
$\entwo$ (keV) & $1.388_{-0.008}^{+0.008}$ & $1.393_{-0.006}^{+0.006}$ & $1.363_{-0.010}^{+0.010}$ & $1.418_{-0.003}^{+0.003}$ & $1.416_{-0.006}^{+0.006}$ \\
$\sigmamtwo$ (keV) & $0.073_{-0.008}^{+0.007}$ & $0.09_{-0.01}^{+0.02}$ & $0.079_{-0.009}^{+0.009}$ & $0.075_{-0.005}^{+0.005}$ & $0.077_{-0.008}^{+0.008}$ \\
$\sigmaptwo$ (keV) & $0.110_{-0.009}^{+0.008}$ & $=\sigmamtwo$ & $0.20_{-0.02}^{+0.03}$ & $=\sigmamtwo$ & $=\sigmamtwo$ \\
$\linedtwo$ (keV) & $0.355_{-0.008}^{+0.007}$ & $0.27_{-0.02}^{+0.02}$ & $0.44_{-0.01}^{+0.01}$ & $0.49_{-0.02}^{+0.02}$ & $0.27_{-0.02}^{+0.02}$ \\
$\enthree$ (keV) & $2.23_{-0.02}^{+0.02}$ &  & $2.1^\ast$ & $2.21_{-0.02}^{+0.02}$ & $2.23_{-0.03}^{+0.03}$ \\
$\sigmamthree$ (keV) & $0.13_{-0.02}^{+0.01}$ &  & $0.1^\ast$ & $0.11_{-0.02}^{+0.02}$ & $0.16_{-0.03}^{+0.03}$ \\
$\sigmapthree$ (keV) & $=\sigmamthree$ &  & $=\sigmamthree$ & $=\sigmamthree$ & $=\sigmamthree$ \\
$\linedthree$ (keV) & $0.19_{-0.02}^{+0.02}$ &  & $0.14_{-0.04}^{+0.04}$ & $0.28_{-0.03}^{+0.04}$ & $0.21_{-0.03}^{+0.03}$ \\
$\chi^2/\mbox{dof}$ & 253.4/77 & 76.9/59 & 120.9/57 & 108.5/61 & 105.0/56 \\
\hline
\multicolumn{6}{l}{$^\ast$ Value is fixed.}
    \end{tabular}
\end{table*}

\section{Results} \label{sec:results}
\subsection{Phase-averaged spectra} \label{sec:resultsavg}
The first spectrum we consider is that obtained by combining all \XMM\ pn observations listed in Table~\ref{tab:xmm_obs}, and this spectrum is shown in Figure~\ref{fig:avgall}.
We fit this spectrum with the two blackbody plus multiple asymmetric absorption line model described in Section~\ref{sec:model}.
Such a fit is effectively a simple test for temporal changes.
Results for a three absorption line model are provided in Table~\ref{tab:avgall} (a two absorption line model gives a worse fit with $\chi^2/\mbox{dof}=331/80=4.1$).
The large reduced $\chi^2$ ($=253/77=3.3$) indicates that a single set of parameters does not adequately describe the combined 2001--2020 spectrum and that there is potentially spectral variations over time.
The fit residuals in the bottom panel of Figure~\ref{fig:avgall} show significant deviations of the model from the data at 0.5--1~keV and at 1.5--2~keV, both of which are due in part to the model failing to capture the shape of the absorption features at these energies.
Keeping in mind the poor fit, the spectral parameters are in qualitative agreement with those found in previous works that use a two blackbody and two (or more) absorption line model like the one used here (\citealt{Bignami2003,DeLuca2004,Mori2005}; see next for more detailed comparisons).
Note that these earlier works derived their emission sizes using a distance of $\approx 2$~kpc from \citet{Giacani2000}, whereas we use a distance of 1.4~kpc from \citet{Eppens2024}.

\begin{figure*}
\centering
\includegraphics[width=0.95\textwidth]{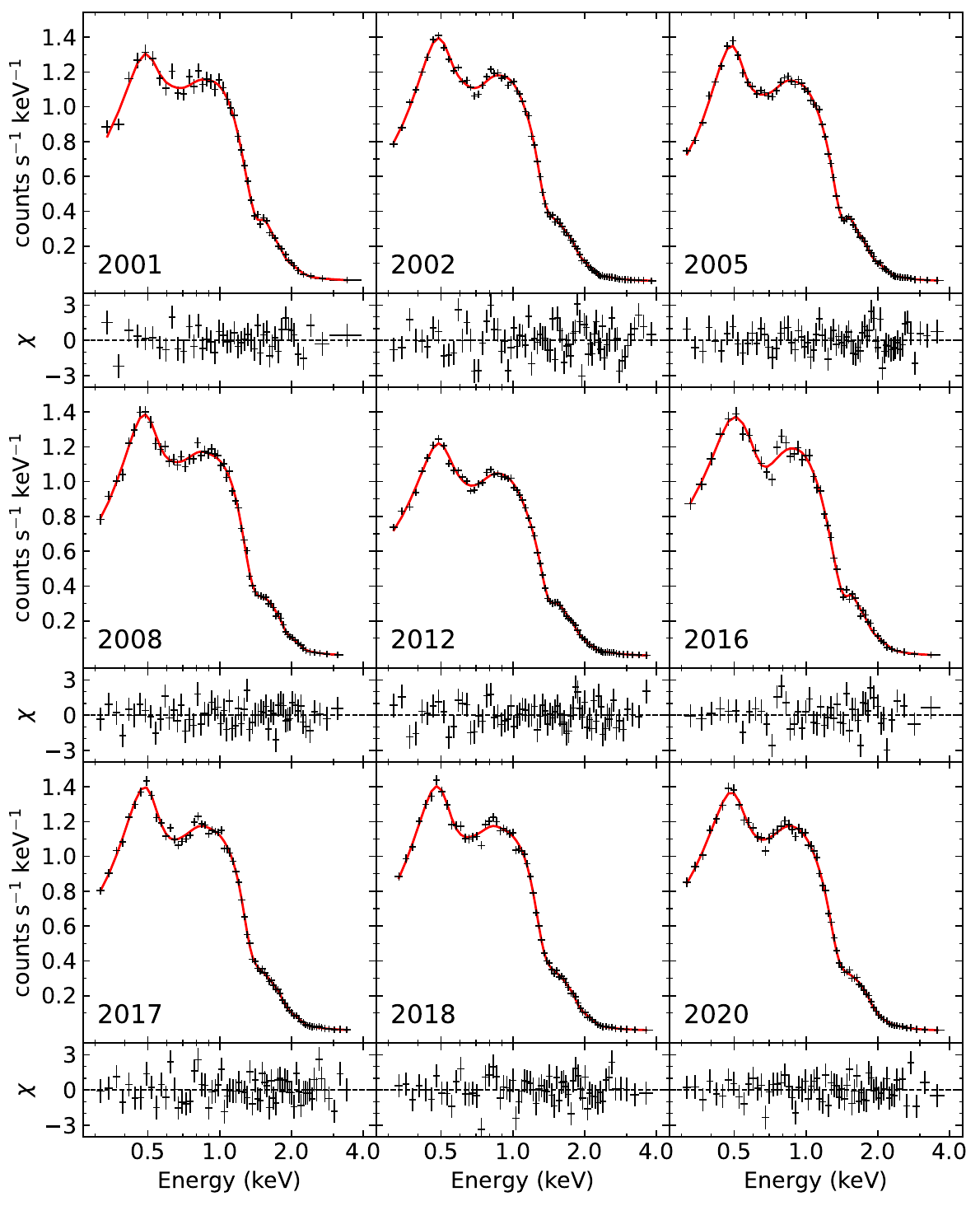}
\caption{\XMM\ EPIC pn spectra of \psr\ at 9 epochs from 2001 to 2020.  Upper panels show data with $1\sigma$ errors (crosses).  Solid lines show spectral model made up of two blackbodies and two or three spectral lines, with best-fitting parameters given in Table~\ref{tab:avg}.  Lower panels show $\chi=(\mbox{data}-\mbox{model})/\mbox{error}$.\label{fig:avg}}
\end{figure*}

\begin{table*}
\renewcommand{\arraystretch}{1.3}
    \centering
    \caption{Phase-averaged spectral modelling results, where the model is composed of two blackbodies and two or three absorption lines and fits are to combined spectra from each epoch.
    $R_1$ and $R_2$ are calculated from the blackbody normalizations assuming a distance $D=1.4\mbox{ kpc}$.
    The absorption line model has a Gaussian shape when line widths are equal, i.e., $\sigmapi=\sigmami$.
    Absorbed 0.3--4~keV flux $f_{0.3-4}^{\rm abs}$ is in units of $10^{-12}\mbox{ erg cm$^{-2}$ s$^{-1}$}$.
    Errors are $1\sigma$.}
    \label{tab:avg}
    \begin{tabular}{|lccccccccr} 
        \hline
        Parameter& 2001 & 2002 & 2005 & 2008 & 2012 & 2016 & 2017 & 2018 & 2020 \\ \hline
$\NH$ ($10^{21}\mbox{ cm$^{-2}$}$) & $0.75_{-0.28}^{+0.35}$ & $0.945_{-0.085}^{+0.097}$ & $1.16_{-0.12}^{+0.21}$ & $0.96_{-0.18}^{+0.23}$ & $0.76_{-0.15}^{+0.18}$ & $0.61_{-0.14}^{+0.17}$ & $0.90_{-0.14}^{+0.14}$ & $0.97_{-0.15}^{+0.30}$ & $0.65_{-0.10}^{+0.15}$ \\
$kT_1$ (keV) & $0.209_{-0.020}^{+0.018}$ & $0.175_{-0.008}^{+0.009}$ & $0.174_{-0.013}^{+0.009}$ & $0.207_{-0.016}^{+0.013}$ & $0.191_{-0.015}^{+0.017}$ & $0.192_{-0.016}^{+0.017}$ & $0.208_{-0.008}^{+0.009}$ & $0.208_{-0.016}^{+0.008}$ & $0.181_{-0.019}^{+0.016}$ \\
$R_1$ (km) & $1.9_{-1.2}^{+1.9}$ & $2.3_{-1.0}^{+1.2}$ & $2.7_{-1.3}^{+2.1}$ & $2.1_{-1.1}^{+1.6}$ & $1.8_{-0.9}^{+1.2}$ & $1.7_{-0.8}^{+1.1}$ & $2.0_{-0.9}^{+1.0}$ & $2.2_{-1.1}^{+2.1}$ & $1.8_{-0.9}^{+1.3}$ \\
$kT_2$ (keV) & $0.41_{-0.05}^{+0.10}$ & $0.318_{-0.007}^{+0.008}$ & $0.321_{-0.011}^{+0.010}$ & $0.393_{-0.054}^{+0.094}$ & $0.342_{-0.017}^{+0.015}$ & $0.330_{-0.017}^{+0.024}$ & $0.404_{-0.030}^{+0.049}$ & $0.442_{-0.039}^{+0.040}$ & $0.311_{-0.013}^{+0.013}$ \\
$R_2$ (km) & $0.17_{-0.14}^{+0.19}$ & $0.52_{-0.23}^{+0.23}$ & $0.47_{-0.23}^{+0.28}$ & $0.19_{-0.17}^{+0.29}$ & $0.36_{-0.26}^{+0.26}$ & $0.44_{-0.29}^{+0.32}$ & $0.18_{-0.13}^{+0.16}$ & $0.12_{-0.08}^{+0.13}$ & $0.55_{-0.30}^{+0.38}$ \\
$\enone$ (keV) & $0.68_{-0.03}^{+0.02}$ & $0.672_{-0.008}^{+0.007}$ & $0.66_{-0.01}^{+0.01}$ & $0.67_{-0.02}^{+0.01}$ & $0.66_{-0.02}^{+0.02}$ & $0.65_{-0.04}^{+0.02}$ & $0.65_{-0.01}^{+0.01}$ & $0.663_{-0.006}^{+0.014}$ & $0.65_{-0.01}^{+0.02}$ \\
$\sigmamone$ (keV) & $0.11_{-0.04}^{+0.06}$ & $0.09_{-0.01}^{+0.01}$ & $0.09_{-0.02}^{+0.03}$ & $0.10_{-0.03}^{+0.04}$ & $0.07_{-0.02}^{+0.02}$ & $0.04_{-0.04}^{+0.03}$ & $0.08_{-0.02}^{+0.02}$ & $0.11_{-0.02}^{+0.05}$ & $0.06_{-0.02}^{+0.02}$ \\
$\sigmapone$ (keV) & $0.29_{-0.05}^{+0.06}$ & $0.23_{-0.02}^{+0.02}$ & $0.27_{-0.03}^{+0.03}$ & $0.35_{-0.05}^{+0.05}$ & $0.23_{-0.03}^{+0.04}$ & $0.19_{-0.03}^{+0.04}$ & $0.35_{-0.03}^{+0.04}$ & $0.40_{-0.03}^{+0.07}$ & $0.23_{-0.03}^{+0.04}$ \\
$\linedone$ (keV) & $0.47_{-0.07}^{+0.10}$ & $0.45_{-0.02}^{+0.02}$ & $0.49_{-0.02}^{+0.04}$ & $0.51_{-0.05}^{+0.07}$ & $0.41_{-0.02}^{+0.03}$ & $0.40_{-0.02}^{+0.02}$ & $0.50_{-0.04}^{+0.04}$ & $0.55_{-0.05}^{+0.11}$ & $0.40_{-0.02}^{+0.03}$ \\
$\entwo$ (keV) & $1.44_{-0.05}^{+0.06}$ & $1.37_{-0.02}^{+0.02}$ & $1.43_{-0.03}^{+0.05}$ & $1.40_{-0.02}^{+0.02}$ & $1.41_{-0.04}^{+0.05}$ & $1.43_{-0.05}^{+0.06}$ & $1.37_{-0.02}^{+0.02}$ & $1.38_{-0.02}^{+0.02}$ & $1.39_{-0.03}^{+0.02}$ \\
$\sigmamtwo$ (keV) & $0.13_{-0.04}^{+0.05}$ & $0.06_{-0.02}^{+0.02}$ & $0.09_{-0.02}^{+0.02}$ & $0.10_{-0.02}^{+0.02}$ & $0.07_{-0.03}^{+0.03}$ & $0.08_{-0.04}^{+0.04}$ & $0.08_{-0.02}^{+0.02}$ & $0.09_{-0.02}^{+0.02}$ & $0.09_{-0.03}^{+0.02}$ \\
$\sigmaptwo$ (keV) & $0.05_{-0.05}^{+0.06}$ & $0.14_{-0.02}^{+0.02}$ & $0.04_{-0.04}^{+0.03}$ & $0.12_{-0.03}^{+0.03}$ & $0.07_{-0.04}^{+0.04}$ & $0.05_{-0.05}^{+0.05}$ & $0.16_{-0.02}^{+0.03}$ & $0.16_{-0.02}^{+0.03}$ & $0.13_{-0.03}^{+0.04}$ \\
$\linedtwo$ (keV) & $0.40_{-0.04}^{+0.07}$ & $0.33_{-0.01}^{+0.01}$ & $0.42_{-0.03}^{+0.08}$ & $0.43_{-0.02}^{+0.02}$ & $0.37_{-0.04}^{+0.05}$ & $0.39_{-0.05}^{+0.07}$ & $0.38_{-0.02}^{+0.02}$ & $0.40_{-0.02}^{+0.02}$ & $0.36_{-0.02}^{+0.02}$ \\
$\enthree$ (keV) &  & $2.22_{-0.03}^{+0.03}$ &  &  & $2.21_{-0.07}^{+0.06}$ & $2.18_{-0.03}^{+0.03}$ &  &  & $2.13_{-0.03}^{+0.03}$ \\
$\sigmamthree$ (keV) & & $0.13_{-0.03}^{+0.03}$ &  &  & $0.13_{-0.08}^{+0.06}$ & $0.04_{-0.04}^{+0.04}$ &  &  & $0.11_{-0.03}^{+0.04}$ \\
$\sigmapthree$ (keV) &  & $=\sigmamthree$ &  &  & $=\sigmamthree$ & $=\sigmamthree$ &  &  & $=\sigmamthree$ \\
$\linedthree$ (keV) &  & $0.20_{-0.03}^{+0.03}$ &  &  & $0.25_{-0.07}^{+0.07}$ & $0.5_{-0.2}^{+2.1}$ &  &  & $0.30_{-0.05}^{+0.05}$ \\
$f_{0.3-4}^{\rm abs}$ &2.23 &2.18 &2.18 &2.14 &1.91 &2.18 &2.15 &2.14 &2.12 \\
$\chi^2/\mbox{dof}$ & 40.1/34 & 141.4/63 & 65.8/60 & 49.3/51 & 74.6/59 & 51.4/36 & 77.3/60 & 70.9/56 & 53.2/52 \\
\hline
    \end{tabular}		
\end{table*}

Next, we fit our model (with two or three absorption lines, depending on fit quality) to the spectrum at each epoch.
The observed and best-fitting model spectra are shown in Figure~\ref{fig:avg}, and
the best-fitting parameter values are provided in Table~\ref{tab:avg}.
Most spectral model fits are satisfactory, with $\chi^2/\mbox{dof}<1.3$.
The exceptions are the one in 2002 with the longest total exposure time
and the one in 2016 with the second shortest exposure time.
Examining more carefully the fit residuals for each epoch, we see that
poorer $\chi^2$ are the result of differences between the data and model
at $0.5-1\mbox{ keV}$ in 2002, 2012, 2016, and 2017,
at $\sim2\mbox{ keV}$ in 2001, 2002, 2005, 2012, and 2016, and
at $2-3\mbox{ keV}$ in 2002, 2017, and 2018.
First, these energy ranges are near the (cyclotron) absorption lines, and the
observed deviations are likely due to a more complicated structure to the lines
than the model we use to describe them here.
Similar deviations are seen in the analysis of the 2002 pn (and MOS) data by
\citet{Mori2005}, although they intentionally did not include in their model
a third spectral line at 2.1~keV since they were testing for its existence.
Spectral model fitting of the 2002 data by \citet{DeLuca2004} also show
deviations at $\sim0.5-0.7\mbox{ keV}$ (in both pn and MOS), even with their different asymmetric absorption line model.
Also of note in relation to the preceding work are the structured residuals just below 2~keV seen in the epochs noted above
at an approximate level of $<9\pm3$~percent, as well as in the combined
spectrum shown in Figure~\ref{fig:avgall} at $<6\pm1.6$~percent and in some
phase-resolved spectral results in Section~\ref{sec:resultsphase}.
\citet{DeLuca2004} found similar residuals of $\sim6$~percent in their analysis
and associated them with a miscalibrated silicon instrumental edge at 1.839~keV.
However, subsequent improvements in the calibration\footnote{\url{https://xmmweb.esac.esa.int/docs/documents/CAL-SRN-0407-1-2.pdf}}
of Al K$\alpha$ at 1.5~keV, with an accuracy that varies at most by
$<1.7$~percent during the period of our observations,
suggest the residuals below 2~keV cannot be fully explained by uncertainties
in the Si edge.  More detailed modelling and use of other data, e.g., from MOS
or \NICER, may provide insights into the origin of the features around 2~keV.

\begin{figure}
\centering
\includegraphics[width=0.95\columnwidth]{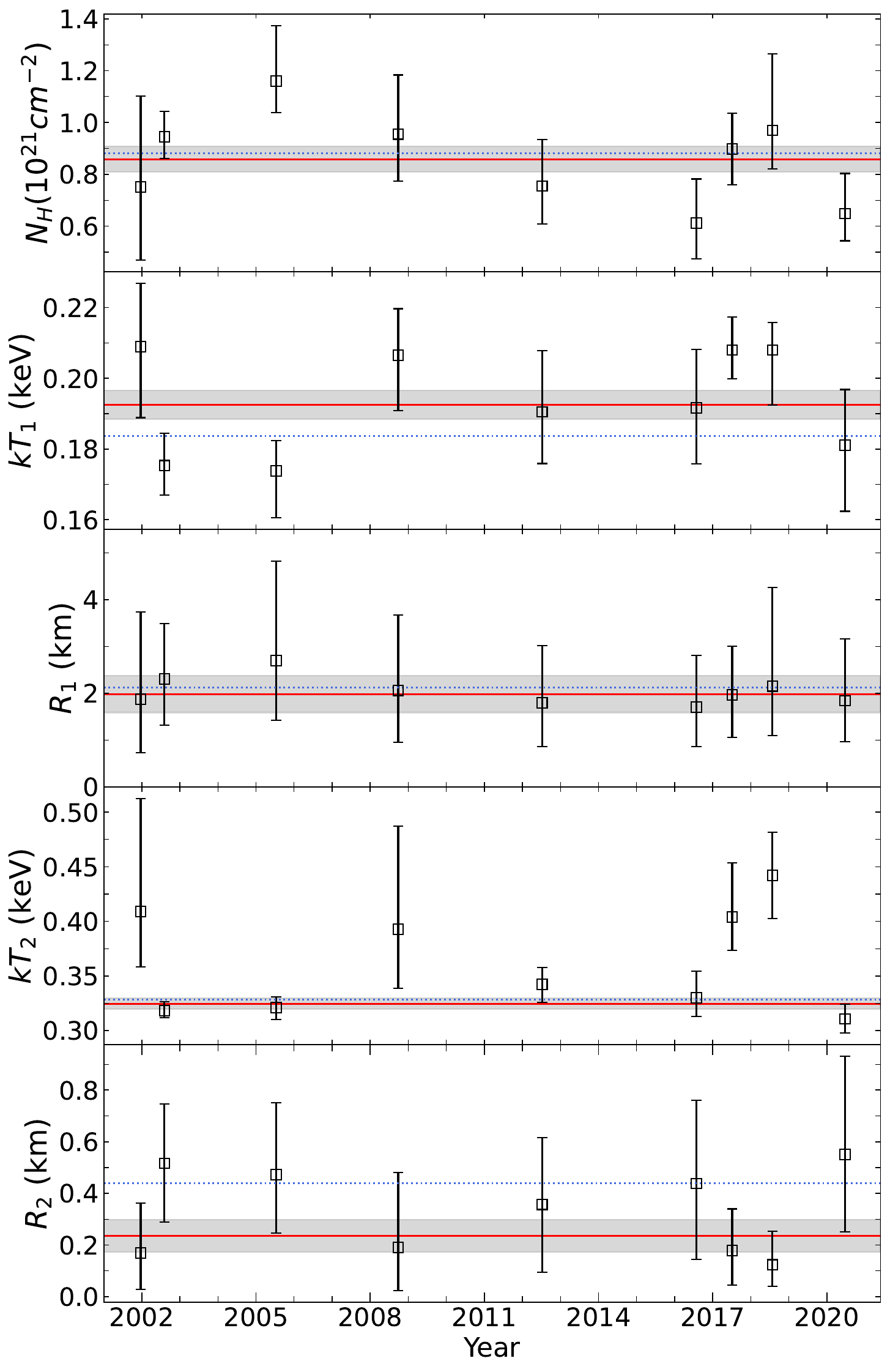}
\caption{Continuum spectral model parameters as a function of time from fits to spectra at 9 epochs (see Table~\ref{tab:avg}).  Panels show interstellar absorption $\NH$ and blackbody emitting region temperatures $kT_1$ and $kT_2$ and radii $R_1$ and $R_2$. Errors are 1$\sigma$. Horizontal solid lines denote a fit of the data to a constant value, with shading indicating $1\sigma$ error of the fit (see Table~\ref{tab:constparamfit}), while dotted lines denote the values from a fit to the spectrum formed from all 9 epochs (see Table~\ref{tab:avgall}).
\label{fig:avgcontin}}
\end{figure}

\begin{figure}
\centering
\includegraphics[width=0.95\columnwidth]{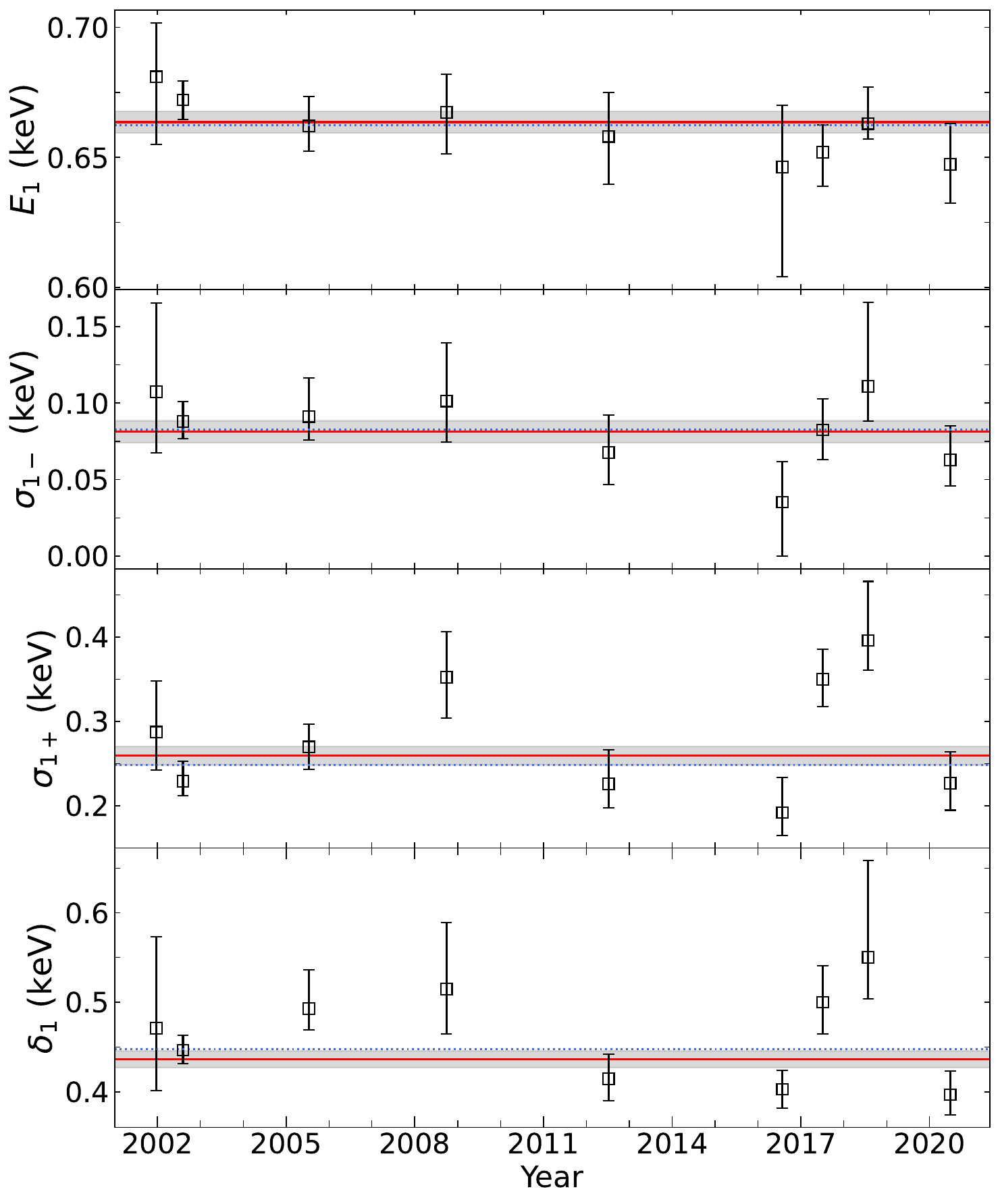}
\caption{Spectral model parameters for first absorption line as a function of time from fits to spectra at 9 epochs (see Table~\ref{tab:avg}).  Panels show line centroid energy $\enone$, widths below and above centroid $\sigmamone$ and $\sigmapone$, and depth $\linedone$. Errors are 1$\sigma$. Horizontal solid lines denote a fit of the data to a constant value, with shading indicating $1\sigma$ error of the fit (see Table~\ref{tab:constparamfit}), while dotted lines denote the values from a fit to the spectrum formed from all 9 epochs (see Table~\ref{tab:avgall}).
\label{fig:avgline1}}
\end{figure}

\begin{figure}
\centering
\includegraphics[width=0.95\columnwidth]{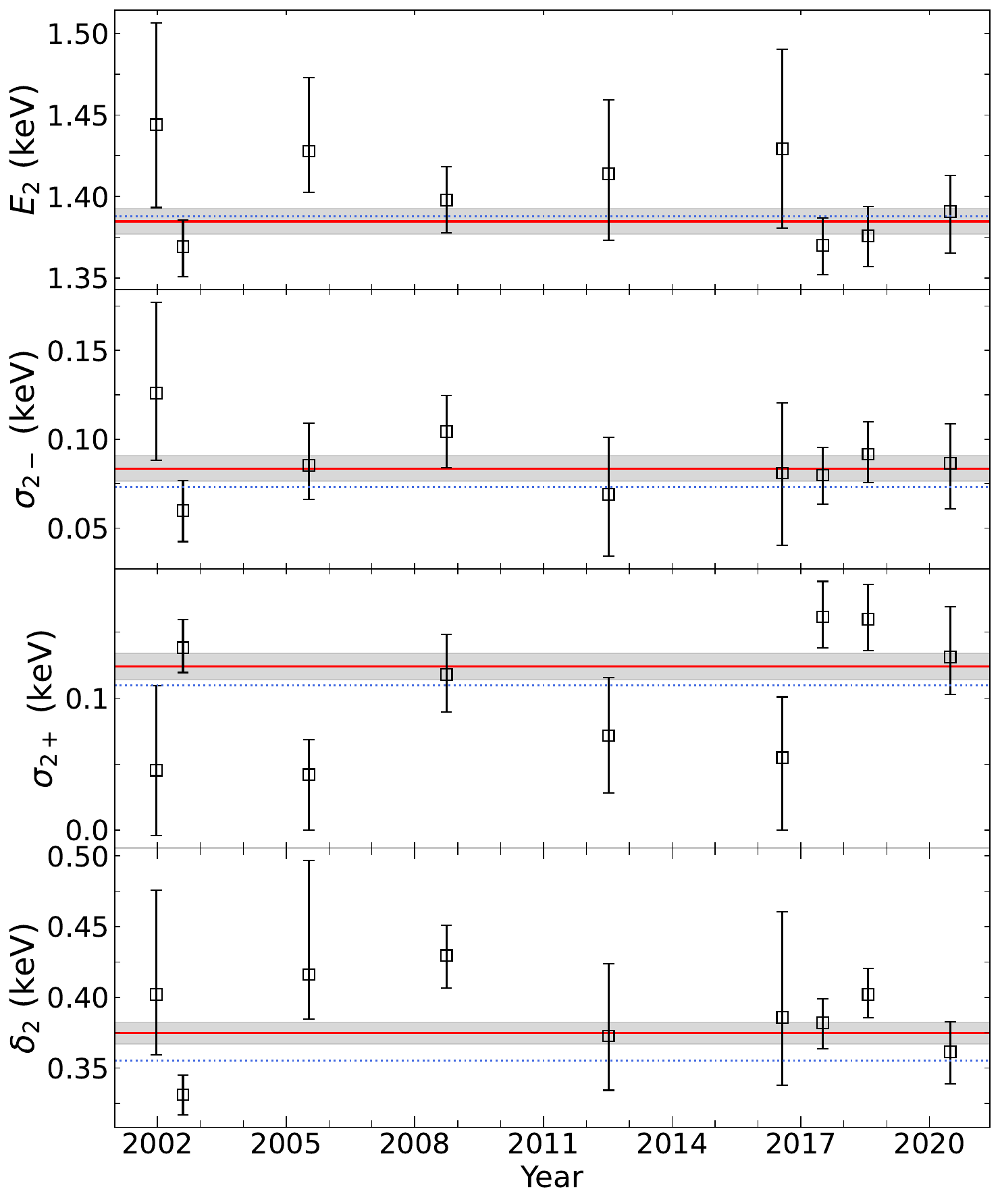}
\caption{Spectral model parameters for second absorption line as a function of time from fits to spectra at 9 epochs (see Table~\ref{tab:avg}).  Panels show line centroid energy $\entwo$, widths below and above centroid $\sigmamtwo$ and $\sigmaptwo$, and depth $\linedtwo$. Errors are 1$\sigma$. Horizontal solid lines denote a fit of the data to a constant value, with shading indicating $1\sigma$ error of the fit (see Table~\ref{tab:constparamfit}), while dotted lines denote the values from a fit to the spectrum formed from all 9 epochs (see Table~\ref{tab:avgall}).
\label{fig:avgline2}}
\end{figure}

\begin{table*}
    \centering
    \caption{Fit of each spectral model parameter to a constant.  For each parameter, the phase-averaged results are obtained by fitting the nine (2001--2020) epoch values from Table~\ref{tab:avg} to a single value over time.  For results at each phase-resolved range (Max, Decline, Min, Rise), seven epochs of values from the corresponding Tables~\ref{tab:max}--\ref{tab:rise} are fit to a single value over time.
The number of degrees of freedom (dof) is 8 and 6 for the phase-averaged and phase-resolved results, respectively.  For phase-resolved data, $\NH$ is not fit to a constant since it is set to each epoch's phase-averaged value.  Errors are $1\sigma$.
\label{tab:constparamfit}}
    \begin{tabular}{lcccccccccc}
        \hline
        Parameter & \multicolumn{2}{c}{Phase-averaged}& \multicolumn{2}{c}{ Max} & \multicolumn{2}{c}{Decline} & \multicolumn{2}{c}{Min} & \multicolumn{2}{c}{Rise} \\
        & Value & $\chi^2/\mbox{dof}$ & Value & $\chi^2/\mbox{dof}$ &Value & $\chi^2/\mbox{dof}$ & Value & $\chi^2/\mbox{dof}$ & Value & $\chi^2/\mbox{dof}$ \\
        \hline
$\NH$ ($10^{21}\mbox{ cm$^{-2}$}$) & 0.86 $\pm$ 0.05 & 1.29 &  &  &  &  &  &  &  &  \\
$kT_1$ (keV) & 0.193 $\pm$ 0.004 & 1.73 & 0.176 $\pm$ 0.004 & 4.55 & 0.194 $\pm$ 0.003 & 5.30 & 0.173 $\pm$ 0.003 & 32.8 & 0.159 $\pm$ 0.004 & 3.83 \\
$R_1$ (km) & 1.98 $\pm$ 0.40 & 0.05 & 2.07 $\pm$ 0.29 & 0.29 & 2.00 $\pm$ 0.29 & 0.41 & 1.93 $\pm$ 0.27 & 0.48 & 2.33 $\pm$ 0.35 & 0.30 \\
$kT_2$ (keV) & 0.325 $\pm$ 0.005 & 2.27 & 0.273 $\pm$ 0.004 & 3.49 & 0.301 $\pm$ 0.013 & 0.62 & 0.262 $\pm$ 0.002 & 0.10 & 0.277 $\pm$ 0.004 & 3.45 \\
$R_2$ (km) & 0.235 $\pm$ 0.062 & 0.67 & 0.47 $\pm$ 0.13 & 0.35 & 0.323 $\pm$ 0.084 & 0.14 & 0.42 $\pm$ 0.084 & 1.54 & 0.61 $\pm$ 0.14 & 0.39 \\
$\enone$ (keV) & 0.664 $\pm$ 0.004 & 0.56 & 0.721 $\pm$ 0.006 & 1.22 & 0.698 $\pm$ 0.006 & 4.28 & 0.666 $\pm$ 0.004 & 5.53 & 0.704 $\pm$ 0.004 & 4.27 \\
$\sigmamone$ (keV) & 0.081 $\pm$ 0.007 & 0.66 & 0.11 $\pm$ 0.01 & 1.41 & 0.13 $\pm$ 0.01 & 4.82 & 0.078 $\pm$ 0.006 & 0.69 & 0.088 $\pm$ 0.006 & 2.57 \\
$\sigmapone$ (keV) & 0.26 $\pm$ 0.01 & 3.18 & 0.16 $\pm$ 0.01 & 2.96 & 0.29 $\pm$ 0.01 & 5.61 & 0.19 $\pm$ 0.01 & 23.5 & 0.12 $\pm$ 0.01 & 2.55 \\
$\linedone$ (keV) & 0.437 $\pm$ 0.009 & 1.99 & 0.37 $\pm$ 0.01 & 3.14 & 0.49 $\pm$ 0.01 & 6.47 & 0.525 $\pm$ 0.008 & 4.15 & 0.442 $\pm$ 0.008 & 1.24 \\
$\entwo$ (keV) & 1.385 $\pm$ 0.008 & 0.74 & 1.396 $\pm$ 0.006 & 0.44 & 1.403 $\pm$ 0.003 & 1.65 & 1.417 $\pm$ 0.003 & 0.22 & 1.415 $\pm$ 0.005 & 0.56 \\
$\sigmamtwo$ (keV) & 0.083 $\pm$ 0.007 & 0.55 & 0.084 $\pm$ 0.009 & 0.35 & 0.093 $\pm$ 0.005 & 1.89 & 0.077 $\pm$ 0.005 & 0.44 & 0.071 $\pm$ 0.007 & 1.27 \\
$\sigmaptwo$ (keV) & 0.12 $\pm$ 0.01 & 1.98 & $=\sigmamtwo$ &  & $=\sigmamtwo$ &  &$=\sigmamtwo$ &  &$=\sigmamtwo$ &  \\
$\linedtwo$ (keV) & 0.375 $\pm$ 0.008 & 2.44 & 0.27 $\pm$ 0.02 & 1.08 & 0.42 $\pm$ 0.01 & 0.63 & 0.51 $\pm$ 0.02 & 0.20 & 0.28 $\pm$ 0.02 & 2.87 \\
        \hline
    \end{tabular}
\end{table*}

Bearing in mind differences in the absorption line models, we find our best-fit parameters for the 2002 data to be consistent
within uncertainties to the pn results of \citet{DeLuca2004}.
Our 2002 parameters also agree qualitatively with those of \citet{Mori2005}.
Meanwhile, our 2001 absorption line parameters are consistent with those of
\citet{Mereghetti2002}, whom modelled the pn spectrum with a single blackbody
and two Gaussian absorption lines.

We plot the spectral parameters of our best-fits in Figures~\ref{fig:avgcontin}, \ref{fig:avgline1}, and \ref{fig:avgline2} for the continuum, first absorption line, and second absorption line components, respectively.
We also show in these figures the corresponding values from
Table~\ref{tab:avgall} for the fit to the combined 2001--2020 spectrum.
In addition, to evaluate whether each spectral model parameter
(e.g., $\NH$ or $kT_1$) varies or is constant over time, we compare their
values at the nine epochs to a single constant value and calculate the
deviations of the comparison (i.e., $\chi^2$ of the best-fit to a constant).
The result of these fits are given in Table~\ref{tab:constparamfit} and
shown in the figures.
Despite variations in the median parameter values, clear changes are not
distinguishable when taking into account the corresponding uncertainties.
The largest departures from constancy are possibly in $kT_1$ and $kT_2$ in 2017 and 2018 for
the continuum parameters, as well $\sigmapone$, $\linedone$, and $\sigmaptwo$ at the same epochs.
The temperature changes in the two components could be correlated, as flux from a lower $kT_2$
is shifted to lower energies and pushes $kT_1$ lower.
Small changes in the derived temperatures could also cause subtle changes in the
accompanying line widths and depths.
Taken as whole, we do not find strong evidence for changes in spectral model
parameters when fitting to the observed phase-averaged spectra of \psr\
from 2001 to 2020.
We consider potential minor temporal variations to more likely be due to uncertainties
in the data and model fitting than to intrinsic changes in the magnetic field.

\subsection{Phase-resolved spectra} \label{sec:resultsphase}
As mentioned in Section~\ref{sec:dataphase}, Figure~\ref{fig:pp} shows the pulse profiles extracted from the \XMM\ pn observations from 2001 to 2020, except for the shorter datasets in 2001 and 2016, and the phase ranges (i.e., maximum, declining, minimum, and rising) used for phase-resolved spectral analyses are displayed in the figure.
We note that the 2001 pulse profile can be found in \citet{Mereghetti2002}, and its phase-resolved spectra, with slightly different phase ranges, are analysed there.
The 2002 pulse profile can be found in \citet{Bignami2003,DeLuca2004}, while the 2001, 2002, 2005, and 2008 pulse profiles are shown in \citet{Halpern2011}.
Kolmogorov-Smirnov (KS) tests indicate no significant differences amongst our (normalized) pulse profiles.
Fitting a sine function to each, we estimate pulsed fractions of 3--6~percent, in approximate agreement with those from the works above.

\begin{figure}
\centering
\includegraphics[width=0.95\columnwidth]{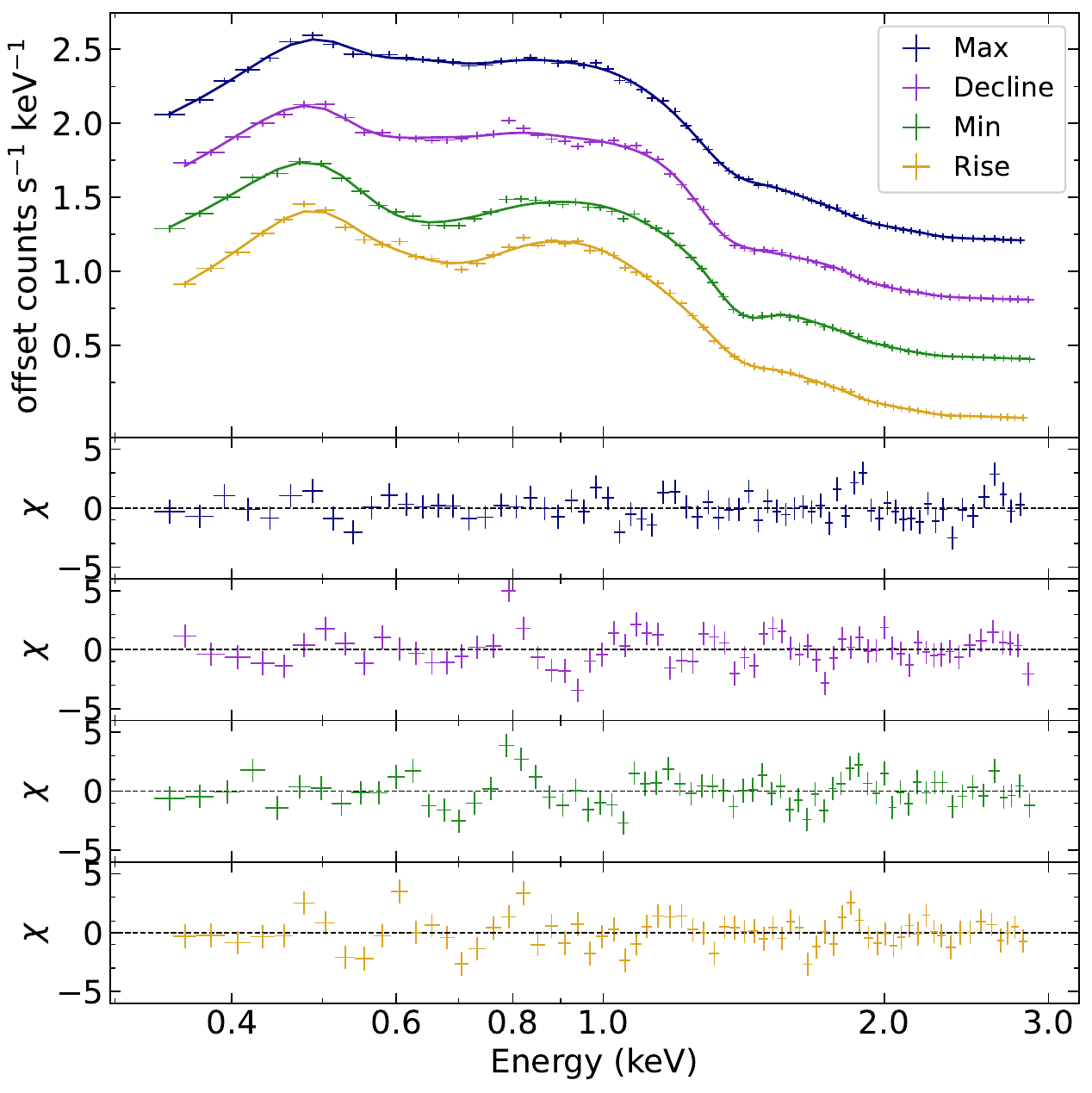}
\caption{\XMM\ EPIC pn spectra of \psr.  Top panel shows data combined from all observations from 2002--2020 and divided into 4 rotation phase ranges (Max, Decline, Min, Rise) with $1\sigma$ errors (crosses).  Solid lines show spectral model made up of two blackbodies and three spectral lines, with best-fitting parameters given in Table~\ref{tab:avgall}.
The Max, Decline, and Min spectra have been offset in the top panel by $+1.2$, $+0.8$, and $+0.4$ counts~s$^{-1}$~keV$^{-1}$, respectively, purely for illustration so that each spectrum does not overlap.
Lower panels show $\chi=(\mbox{data}-\mbox{model})/\mbox{error}$.
\label{fig:all}}
\end{figure}

\begin{figure*}
\centering
\includegraphics[width=0.95\textwidth]{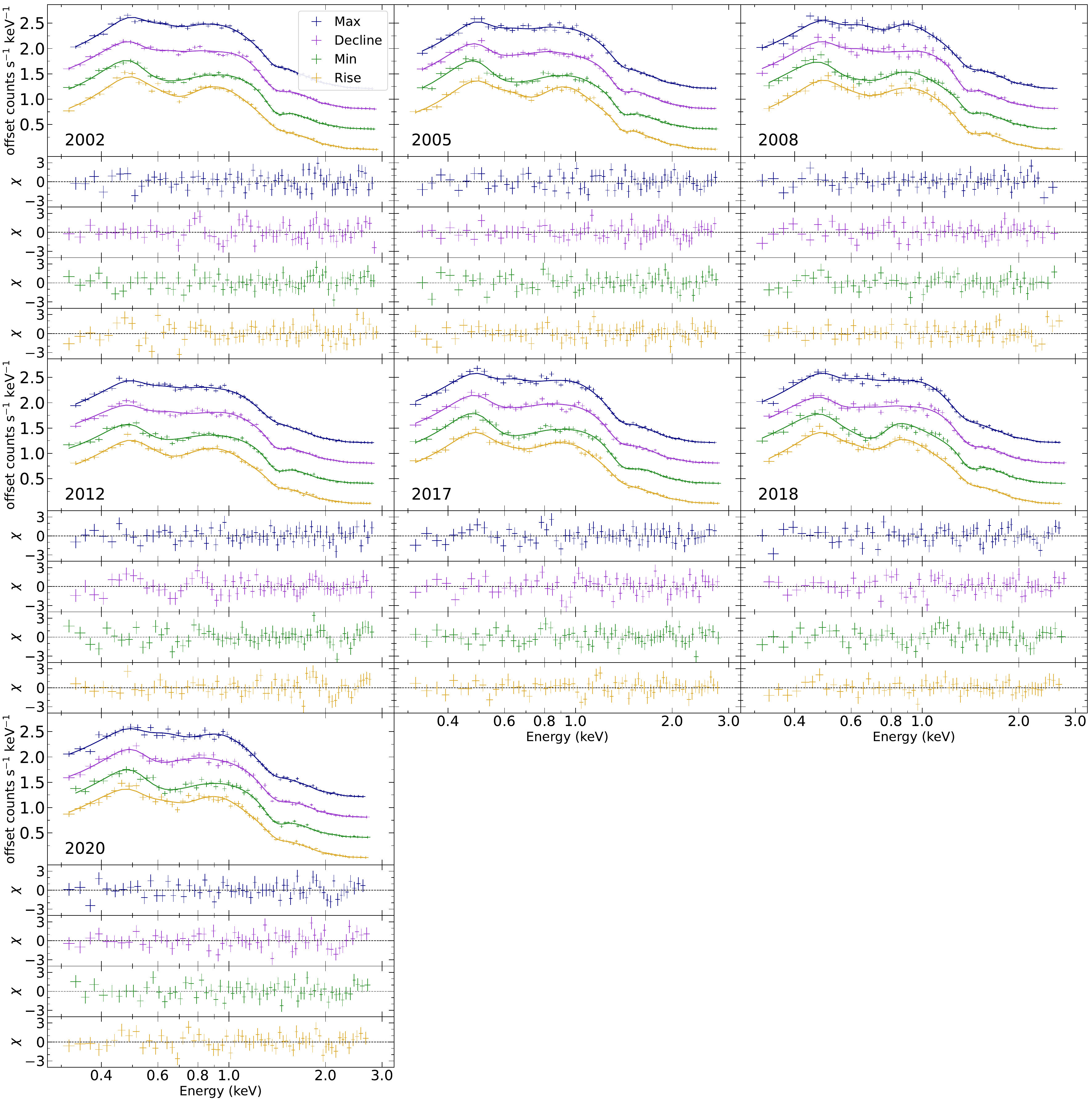}
\caption{\XMM\ EPIC pn spectra of \psr\ at 7 epochs from 2002 to 2020.  Upper panels show data divided into 4 rotation phase ranges (Max, Decline, Min, Rise) with $1\sigma$ errors (crosses).  Solid lines show spectral model made up of two blackbodies and two spectral lines, with best-fitting parameters given in Tables~\ref{tab:max}--\ref{tab:rise}.
The Max, Decline, and Min spectra have been offset in the top panels by $+1.2$, $+0.8$, and $+0.4$ counts~s$^{-1}$~keV$^{-1}$, respectively, purely for illustration so that each spectrum does not overlap.
Lower panels show $\chi=(\mbox{data}-\mbox{model})/\mbox{error}$.
\label{fig:each}}
\end{figure*}

As with the phase--averaged analysis, we first test whether model parameters derived from the phase-resolved spectra are constant throughout the span of data, and this is done by combining and fitting all observations simultaneously. The combined phase-resolved spectrum for each phase range is shown in Figure~\ref{fig:all}, and the best-fitting results are given in Tables~\ref{tab:avgall}. We find the goodness-of-fit for each phase range, except the maximum phase, is $\chi^2/\mbox{dof}\approx1.8-2.1$, indicating the phase-resolved spectra are inconsistent with spectral model parameters that are constant over the approximately 20-year span of data.
Therefore, we perform spectral model fits for each individual epoch, with results shown in Figure~\ref{fig:each} and best-fitting parameters presented in Tables~\ref{tab:max}--\ref{tab:rise}.
Due to the total exposure time contributing to each spectrum being reduced (i.e., divided into four phases), we limit our spectral model to two absorption lines.

\begin{table*}
\renewcommand{\arraystretch}{1.3}
    \centering
    \caption{Phase-resolved spectral modelling results for each epoch (phase range: Maximum), where the model is composed of two blackbodies and two absorption lines.
$\NH$ is fixed at the best-fit phase-averaged value obtained for each epoch (see Table~\ref{tab:avg}).
    $R_1$ and $R_2$ are calculated from the blackbody normalizations assuming a distance $D=1.4\mbox{ kpc}$.
    The absorption line model has a Gaussian shape when line widths are equal, i.e., $\sigmapi=\sigmami$.
    Errors are $1\sigma$.}
    \label{tab:max}
    \begin{tabular}{|lccccccccr} 
        \hline
        Parameter& 2002 & 2005 & 2008 & 2012 & 2017 & 2018 & 2020 \\ \hline
$\NH$ ($10^{21}\mbox{ cm$^{-2}$}$) & 0.945&  1.16&  0.96&  0.76&  0.90&    0.97& 0.65\\
$kT_1$ (keV) & $0.176_{-0.009}^{+0.009}$ & $0.175_{-0.012}^{+0.016}$ & $0.108_{-0.014}^{+0.018}$ & $0.204_{-0.014}^{+0.004}$ & $0.170_{-0.011}^{+0.010}$ & $0.174_{-0.009}^{+0.009}$ & $0.166_{-0.027}^{+0.020}$ \\
$R_1$ (km) & $2.17_{-0.64}^{+0.68}$ & $2.5_{-1.0}^{+1.1}$ & $4.1_{-2.6}^{+4.5}$ & $1.69_{-0.43}^{+0.53}$ & $2.45_{-0.79}^{+0.91}$ & $2.62_{-0.91}^{+0.97}$ & $1.9_{-0.7}^{+1.0}$ \\
$kT_2$ (keV) & $0.304_{-0.012}^{+0.015}$ & $0.311_{-0.018}^{+0.037}$ & $0.262_{-0.004}^{+0.006}$ & $0.350_{-0.033}^\ast$ & $0.303_{-0.017}^{+0.023}$ & $0.322_{-0.023}^{+0.032}$ & $0.284_{-0.016}^{+0.027}$ \\
$R_2$ (km) & $0.59_{-0.33}^{+0.36}$ & $0.54_{-0.41}^{+0.41}$ & $1.04_{-0.42}^{+0.85}$ & $0.29_{-0.06}^{+0.34}$ & $0.59_{-0.38}^{+0.44}$ & $0.46_{-0.32}^{+0.32}$ & $0.75_{-0.53}^{+0.57}$ \\
$\enone$ (keV) & $0.69_{-0.02}^{+0.02}$ & $0.70_{-0.02}^{+0.02}$ & $0.77_{-0.05}^{+0.07}$ & $0.70_{-0.02}^{+0.02}$ & $0.73_{-0.01}^{+0.01}$ & $0.73_{-0.02}^{+0.02}$ & $0.74_{-0.01}^{+0.01}$ \\
$\sigmamone$ (keV) & $0.09_{-0.02}^{+0.02}$ & $0.11_{-0.03}^{+0.04}$ & $0.12_{-0.07}^{+0.08}$ & $0.11_{-0.01}^{+0.03}$ & $0.18_{-0.02}^{+0.04}$ & $0.20_{-0.02}^{+0.00}$ & $0.12_{-0.02}^{+0.02}$ \\
$\sigmapone$ (keV) & $0.19_{-0.02}^{+0.03}$ & $0.24_{-0.04}^{+0.06}$ & $0.06_{-0.06}^{+0.04}$ & $0.24_{-0.05}^{+0.03}$ & $=\sigmamone$ & $=\sigmamone$ & $=\sigmamone$ \\
$\linedone$ (keV) & $0.36_{-0.02}^{+0.02}$ & $0.41_{-0.03}^{+0.04}$ & $0.26_{-0.05}^{+0.04}$ & $0.37_{-0.02}^{+0.03}$ & $0.40_{-0.03}^{+0.05}$ & $0.44_{-0.03}^{+0.02}$ & $0.32_{-0.04}^{+0.03}$ \\
$\entwo$ (keV) & $1.39_{-0.01}^{+0.01}$ & $1.41_{-0.01}^{+0.01}$ & $1.40_{-0.02}^{+0.02}$ & $1.39_{-0.02}^{+0.02}$ & $1.40_{-0.01}^{+0.01}$ & $1.37_{-0.03}^{+0.02}$ & $1.39_{-0.02}^{+0.02}$ \\
$\sigmamtwo$ (keV) & $0.07_{-0.02}^{+0.02}$ & $0.08_{-0.02}^{+0.02}$ & $0.09_{-0.02}^{+0.03}$ & $0.11_{-0.03}^{+0.03}$ & $0.08_{-0.02}^{+0.02}$ & $0.10_{-0.03}^{+0.03}$ & $0.09_{-0.02}^{+0.03}$ \\
$\sigmaptwo$ (keV) & $=\sigmamtwo$ & $=\sigmamtwo$ & $=\sigmamtwo$ & $=\sigmamtwo$ & $=\sigmamtwo$ & $=\sigmamtwo$ & $=\sigmamtwo$ \\
$\linedtwo$ (keV) & $0.27_{-0.03}^{+0.04}$ & $0.28_{-0.04}^{+0.04}$ & $0.33_{-0.05}^{+0.05}$ & $0.22_{-0.03}^{+0.03}$ & $0.32_{-0.04}^{+0.04}$ & $0.24_{-0.04}^{+0.04}$ & $0.28_{-0.04}^{+0.05}$ \\
$\chi^2/\mbox{dof}$ & 83.7/59 & 71.7/56 & 56.3/50 & 62.4/57 & 63.9/57 & 75.6/56 & 67.0/54 \\
\hline
\multicolumn{8}{l}{$^\ast$ Value pegged at maximum.}
    \end{tabular}
\end{table*}

\begin{table*}
\renewcommand{\arraystretch}{1.3}
    \centering
    \caption{Phase-resolved spectral modelling results for each epoch (phase range: Declining), where the model is composed of two blackbodies and two absorption lines.
$\NH$ is fixed at the best-fit phase-averaged value obtained for each epoch (see Table~\ref{tab:avg}).
    $R_1$ and $R_2$ are calculated from the blackbody normalizations assuming a distance $D=1.4\mbox{ kpc}$.
    The absorption line model has a Gaussian shape when line widths are equal, i.e., $\sigmapi=\sigmami$.
    Errors are $1\sigma$.}
    \label{tab:decline}
    \begin{tabular}{|lccccccccr} 
        \hline
        Parameter& 2002 & 2005 & 2008 & 2012 & 2017 & 2018 & 2020 \\ \hline
$\NH$ ($10^{21}\mbox{ cm$^{-2}$}$) & 0.945&  1.16&  0.96&  0.76&  0.90&    0.97& 0.65\\
$kT_1$ (keV) & $0.191_{-0.012}^{+0.011}$ & $0.188_{-0.008}^{+0.004}$ & $0.178_{-0.010}^{+0.010}$ & $0.168_{-0.012}^{+0.010}$ & $0.200_{-0.006}^{+0.004}$ & $0.182_{-0.005}^{+0.007}$ & $0.219_{-0.007}^{+0.005}$ \\
$R_1$ (km) & $2.3_{-0.9}^{+1.1}$ & $2.7_{-1.0}^{+1.4}$ & $2.6_{-1.0}^{+1.2}$ & $2.18_{-0.73}^{+0.85}$ & $2.05_{-0.65}^{+0.82}$ & $3.2_{-1.7}^{+1.2}$ & $1.57_{-0.42}^{+0.46}$ \\
$kT_2$ (keV) & $0.328_{-0.026}^{+0.048}$ & $0.350_{-0.025}^\ast$ & $0.328_{-0.031}^{+0.051}$ & $0.293_{-0.014}^{+0.017}$ & $0.350_{-0.019}^\ast$ & $0.350_{-0.023}^\ast$ & $0.350_{-0.040}^\ast$ \\
$R_2$ (km) & $0.40_{-0.33}^{+0.41}$ & $0.32_{-0.06}^{+0.29}$ & $0.41_{-0.30}^{+0.38}$ & $0.61_{-0.36}^{+0.41}$ & $0.29_{-0.08}^{+0.22}$ & $0.30_{-0.10}^{+0.31}$ & $0.25_{-0.14}^{+0.34}$ \\
$\enone$ (keV) & $0.71_{-0.01}^{+0.01}$ & $0.66_{-0.02}^{+0.02}$ & $0.74_{-0.02}^{+0.02}$ & $0.72_{-0.01}^{+0.01}$ & $0.66_{-0.02}^{+0.02}$ & $0.68_{-0.02}^{+0.02}$ & $0.62_{-0.04}^{+0.03}$ \\
$\sigmamone$ (keV) & $0.18_{-0.03}^{+0.05}$ & $0.13_{-0.03}^{+0.05}$ & $0.21_{-0.03}^{+0.04}$ & $0.19_{-0.02}^{+0.03}$ & $0.10_{-0.03}^{+0.02}$ & $0.25_{-0.06}^\ast$ & $0.05_{-0.02}^{+0.03}$ \\
$\sigmapone$ (keV) & $0.32_{-0.03}^{+0.04}$ & $0.40_{-0.04}^{+0.04}$ & $=\sigmamone$ & $=\sigmamone$ & $0.32_{-0.03}^{+0.03}$ & $0.39_{-0.04}^{+0.04}$ & $0.36_{-0.05}^{+0.05}$ \\
$\linedone$ (keV) & $0.54_{-0.04}^{+0.05}$ & $0.56_{-0.04}^{+0.03}$ & $0.51_{-0.04}^{+0.06}$ & $0.44_{-0.04}^{+0.04}$ & $0.49_{-0.03}^{+0.02}$ & $0.67_{-0.06}^{+0.03}$ & $0.41_{-0.02}^{+0.02}$ \\
$\entwo$ (keV) & $1.402_{-0.007}^{+0.008}$ & $1.399_{-0.007}^{+0.008}$ & $1.37_{-0.01}^{+0.01}$ & $1.413_{-0.008}^{+0.008}$ & $1.39_{-0.01}^{+0.01}$ & $1.41_{-0.01}^{+0.01}$ & $1.41_{-0.01}^{+0.01}$ \\
$\sigmamtwo$ (keV) & $0.11_{-0.01}^{+0.01}$ & $0.08_{-0.01}^{+0.01}$ & $0.07_{-0.02}^{+0.03}$ & $0.07_{-0.01}^{+0.01}$ & $0.10_{-0.01}^{+0.01}$ & $0.10_{-0.01}^{+0.01}$ & $0.11_{-0.01}^{+0.02}$ \\
$\sigmaptwo$ (keV) & $=\sigmamtwo$ & $=\sigmamtwo$ & $=\sigmamtwo$ & $=\sigmamtwo$ & $=\sigmamtwo$ & $=\sigmamtwo$ & $=\sigmamtwo$ \\
$\linedtwo$ (keV) & $0.43_{-0.02}^{+0.02}$ & $0.46_{-0.04}^{+0.04}$ & $0.46_{-0.06}^{+0.10}$ & $0.46_{-0.04}^{+0.06}$ & $0.38_{-0.03}^{+0.03}$ & $0.43_{-0.03}^{+0.03}$ & $0.43_{-0.03}^{+0.03}$ \\
$\chi^2/\mbox{dof}$ & 91.8/59 & 57.3/57 & 56.6/52 & 75.5/58 & 91.9/57 & 79.3/55 & 82.2/54 \\
\hline
\multicolumn{8}{l}{$^\ast$ Value pegged at maximum.}
    \end{tabular}
\end{table*}

\begin{table*}
\renewcommand{\arraystretch}{1.3}
    \centering
    \caption{Phase-resolved spectral modelling results for each epoch (phase range: Minimum), where the model is composed of two blackbodies and two absorption lines.
$\NH$ is fixed at the best-fit phase-averaged value obtained for each epoch (see Table~\ref{tab:avg}).
    $R_1$ and $R_2$ are calculated from the blackbody normalizations assuming a distance $D=1.4\mbox{ kpc}$.
    The absorption line model has a Gaussian shape when line widths are equal, i.e., $\sigmapi=\sigmami$.
    Errors are $1\sigma$.}
    \label{tab:min}
    \begin{tabular}{|lccccccccr} 
        \hline
        Parameter& 2002 & 2005 & 2008 & 2012 & 2017 & 2018 & 2020 \\ \hline
$\NH$ ($10^{21}\mbox{ cm$^{-2}$}$) & 0.945&  1.16&  0.96&  0.76&  0.90&    0.97& 0.65\\
$kT_1$ (keV) & $0.189_{-0.007}^{+0.003}$ & $0.158_{-0.014}^{+0.027}$ & $0.113_{-0.011}^{+0.025}$ & $0.203_{-0.011}^{+0.003}$ & $0.200_{-0.045}^{+0.005}$ & $0.089_{-0.006}^{+0.008}$ & $0.202_{-0.009}^{+0.006}$ \\
$R_1$ (km) & $2.26_{-0.59}^{+0.50}$ & $3.2_{-1.8}^{+1.7}$ & $5_{-2}^{+13}$ & $1.71_{-0.45}^{+0.38}$ & $2.1_{-0.6}^{+1.4}$ & $7.5_{-4.3}^{+5.1}$ & $1.75_{-0.51}^{+0.58}$ \\
$kT_2$ (keV) & $0.350_{-0.019}^{\ast}$ & $0.301_{-0.017}^{+0.049}$ & $0.266_{-0.006}^{+0.019}$ & $0.350_{-0.031}^{\ast}$ & $0.350_{-0.075}^{\ast}$ & $0.262_{-0.002}^{+0.003}$ & $0.350_{-0.026}^{\ast}$ \\
$R_2$ (km) & $0.33_{-0.05}^{+0.24}$ & $0.59_{-0.49}^{+0.45}$ & $0.94_{-0.59}^{+0.38}$ & $0.28_{-0.06}^{+0.30}$ & $0.29_{-0.09}^{+0.62}$ & $1.01_{-0.23}^{+0.23}$ & $0.30_{-0.10}^{+0.27}$ \\
$\enone$ (keV) & $0.653_{-0.005}^{+0.009}$ & $0.65_{-0.01}^{+0.01}$ & $0.68_{-0.01}^{+0.01}$ & $0.64_{-0.02}^{+0.01}$ & $0.64_{-0.01}^{+0.01}$ & $0.690_{-0.006}^{+0.006}$ & $0.64_{-0.02}^{+0.02}$ \\
$\sigmamone$ (keV) & $0.09_{-0.01}^{+0.01}$ & $0.09_{-0.02}^{+0.02}$ & $0.13_{-0.03}^{+0.06}$ & $0.07_{-0.01}^{+0.02}$ & $0.07_{-0.02}^{+0.02}$ & $0.07_{-0.01}^{+0.02}$ & $0.06_{-0.02}^{+0.02}$ \\
$\sigmapone$ (keV) & $0.29_{-0.01}^{+0.02}$ & $0.26_{-0.04}^{+0.13}$ & $=\sigmamone$ & $0.30_{-0.04}^{+0.03}$ & $0.35_{-0.11}^{+0.02}$ & $=\sigmamone$ & $0.27_{-0.03}^{+0.03}$ \\
$\linedone$ (keV) & $0.56_{-0.01}^{+0.01}$ & $0.57_{-0.03}^{+0.06}$ & $0.44_{-0.06}^{+0.12}$ & $0.49_{-0.02}^{+0.02}$ & $0.58_{-0.06}^{+0.02}$ & $0.43_{-0.03}^{+0.03}$ & $0.50_{-0.02}^{+0.02}$ \\
$\entwo$ (keV) & $1.417_{-0.005}^{+0.005}$ & $1.417_{-0.007}^{+0.008}$ & $1.41_{-0.01}^{+0.01}$ & $1.417_{-0.006}^{+0.006}$ & $1.426_{-0.009}^{+0.009}$ & $1.418_{-0.008}^{+0.008}$ & $1.413_{-0.008}^{+0.008}$ \\
$\sigmamtwo$ (keV) & $0.08_{-0.01}^{+0.01}$ & $0.06_{-0.01}^{+0.03}$ & $0.07_{-0.02}^{+0.02}$ & $0.08_{-0.01}^{+0.01}$ & $0.10_{-0.02}^{+0.01}$ & $0.07_{-0.01}^{+0.01}$ & $0.08_{-0.01}^{+0.01}$ \\
$\sigmaptwo$ (keV) & $=\sigmamtwo$ & $=\sigmamtwo$ & $=\sigmamtwo$ & $=\sigmamtwo$ & $=\sigmamtwo$ & $=\sigmamtwo$ & $=\sigmamtwo$ \\
$\linedtwo$ (keV) & $0.51_{-0.03}^{+0.03}$ & $0.50_{-0.05}^{+0.06}$ & $0.45_{-0.07}^{+0.09}$ & $0.53_{-0.04}^{+0.04}$ & $0.49_{-0.05}^{+0.03}$ & $0.52_{-0.04}^{+0.05}$ & $0.52_{-0.04}^{+0.04}$ \\
$\chi^2/\mbox{dof}$ & 96.4/60 & 76.2/57 & 53.2/51 & 88.8/58 & 64.1/57 & 73.2/55 & 67.5/53 \\
\hline
\multicolumn{8}{l}{$^\ast$ Value pegged at maximum.}
    \end{tabular}
\end{table*}

\begin{table*}
\renewcommand{\arraystretch}{1.3}
    \centering
    \caption{Phase-resolved spectral modelling results for each epoch (phase range: Rising), where the model is composed of two blackbodies and two absorption lines.
$\NH$ is fixed at the best-fit phase-averaged value obtained for each epoch (see Table~\ref{tab:avg}).
    $R_1$ and $R_2$ are calculated from the blackbody normalizations assuming a distance $D=1.4\mbox{ kpc}$.
    The absorption line model has a Gaussian shape when line widths are equal, i.e., $\sigmapi=\sigmami$.
    Errors are $1\sigma$.}
    \label{tab:rise}
    \begin{tabular}{|lccccccccr} 
        \hline
        Parameter& 2002 & 2005 & 2008 & 2012 & 2017 & 2018 & 2020 \\ \hline
$\NH$ ($10^{21}\mbox{ cm$^{-2}$}$) & 0.945&  1.16&  0.96&  0.76&  0.90&    0.97& 0.65\\
$kT_1$ (keV) & $0.154_{-0.009}^{+0.008}$ & $0.148_{-0.008}^{+0.006}$ & $0.138_{-0.033}^{+0.028}$ & $0.181_{-0.011}^{+0.009}$ & $0.179_{-0.008}^{+0.009}$ & $0.117_{-0.013}^{+0.015}$ & $0.156_{-0.023}^{+0.014}$ \\
$R_1$ (km) & $2.63_{-0.82}^{+0.99}$ & $3.2_{-1.0}^{+1.2}$ & $2.7_{-1.6}^{+2.5}$ & $1.89_{-0.52}^{+0.59}$ & $2.31_{-0.73}^{+0.80}$ & $3.9_{-2.1}^{+2.9}$ & $2.3_{-0.8}^{+1.2}$ \\
$kT_2$ (keV) & $0.290_{-0.009}^{+0.010}$ & $0.289_{-0.010}^{+0.011}$ & $0.266_{-0.008}^{+0.019}$ & $0.317_{-0.021}^{+0.027}$ & $0.344_{-0.022}^{+0.029}$ & $0.261_{-0.005}^{+0.007}$ & $0.290_{-0.018}^{+0.024}$ \\
$R_2$ (km) & $0.71_{-0.33}^{+0.37}$ & $0.72_{-0.35}^{+0.39}$ & $0.95_{-0.63}^{+0.90}$ & $0.45_{-0.30}^{+0.37}$ & $0.38_{-0.24}^{+0.28}$ & $1.00_{-0.43}^{+0.37}$ & $0.69_{-0.44}^{+0.52}$ \\
$\enone$ (keV) & $0.68_{-0.01}^{+0.01}$ & $0.723_{-0.006}^{+0.006}$ & $0.67_{-0.04}^{+0.03}$ & $0.65_{-0.02}^{+0.02}$ & $0.68_{-0.02}^{+0.02}$ & $0.703_{-0.007}^{+0.007}$ & $0.701_{-0.010}^{+0.009}$ \\
$\sigmamone$ (keV) & $0.07_{-0.01}^{+0.01}$ & $0.11_{-0.01}^{+0.01}$ & $0.07_{-0.04}^{+0.03}$ & $0.05_{-0.02}^{+0.02}$ & $0.09_{-0.02}^{+0.02}$ & $0.09_{-0.01}^{+0.02}$ & $0.13_{-0.02}^{+0.02}$ \\
$\sigmapone$ (keV) & $0.14_{-0.02}^{+0.02}$ & $=\sigmamone$ & $0.16_{-0.04}^{+0.05}$ & $0.17_{-0.02}^{+0.03}$ & $0.18_{-0.03}^{+0.03}$ & $=\sigmamone$ & $=\sigmamone$ \\
$\linedone$ (keV) & $0.45_{-0.02}^{+0.01}$ & $0.45_{-0.02}^{+0.02}$ & $0.40_{-0.06}^{+0.04}$ & $0.44_{-0.02}^{+0.02}$ & $0.46_{-0.02}^{+0.02}$ & $0.37_{-0.04}^{+0.03}$ & $0.40_{-0.04}^{+0.03}$ \\
$\entwo$ (keV) & $1.43_{-0.01}^{+0.01}$ & $1.404_{-0.009}^{+0.009}$ & $1.42_{-0.01}^{+0.01}$ & $1.42_{-0.01}^{+0.01}$ & $1.42_{-0.02}^{+0.02}$ & $1.42_{-0.02}^{+0.02}$ & $1.42_{-0.02}^{+0.02}$ \\
$\sigmamtwo$ (keV) & $0.10_{-0.02}^{+0.02}$ & $0.05_{-0.01}^{+0.01}$ & $0.08_{-0.02}^{+0.02}$ & $0.06_{-0.02}^{+0.02}$ & $0.08_{-0.02}^{+0.03}$ & $0.09_{-0.02}^{+0.02}$ & $0.08_{-0.02}^{+0.02}$ \\
$\sigmaptwo$ (keV) & $=\sigmamtwo$ & $=\sigmamtwo$ & $=\sigmamtwo$ & $=\sigmamtwo$ & $=\sigmamtwo$ & $=\sigmamtwo$ & $=\sigmamtwo$ \\
$\linedtwo$ (keV) & $0.23_{-0.03}^{+0.03}$ & $0.43_{-0.08}^{+0.14}$ & $0.45_{-0.05}^{+0.06}$ & $0.31_{-0.05}^{+0.08}$ & $0.22_{-0.04}^{+0.04}$ & $0.29_{-0.04}^{+0.04}$ & $0.29_{-0.05}^{+0.06}$ \\
$\chi^2/\mbox{dof}$ & 115.7/61 & 60.2/58 & 61.0/50 & 84.0/57 & 58.8/57 & 44.8/55 & 66.4/55 \\
\hline
    \end{tabular}
\end{table*}

\begin{figure*}
\centering
\includegraphics[width=0.95\textwidth]{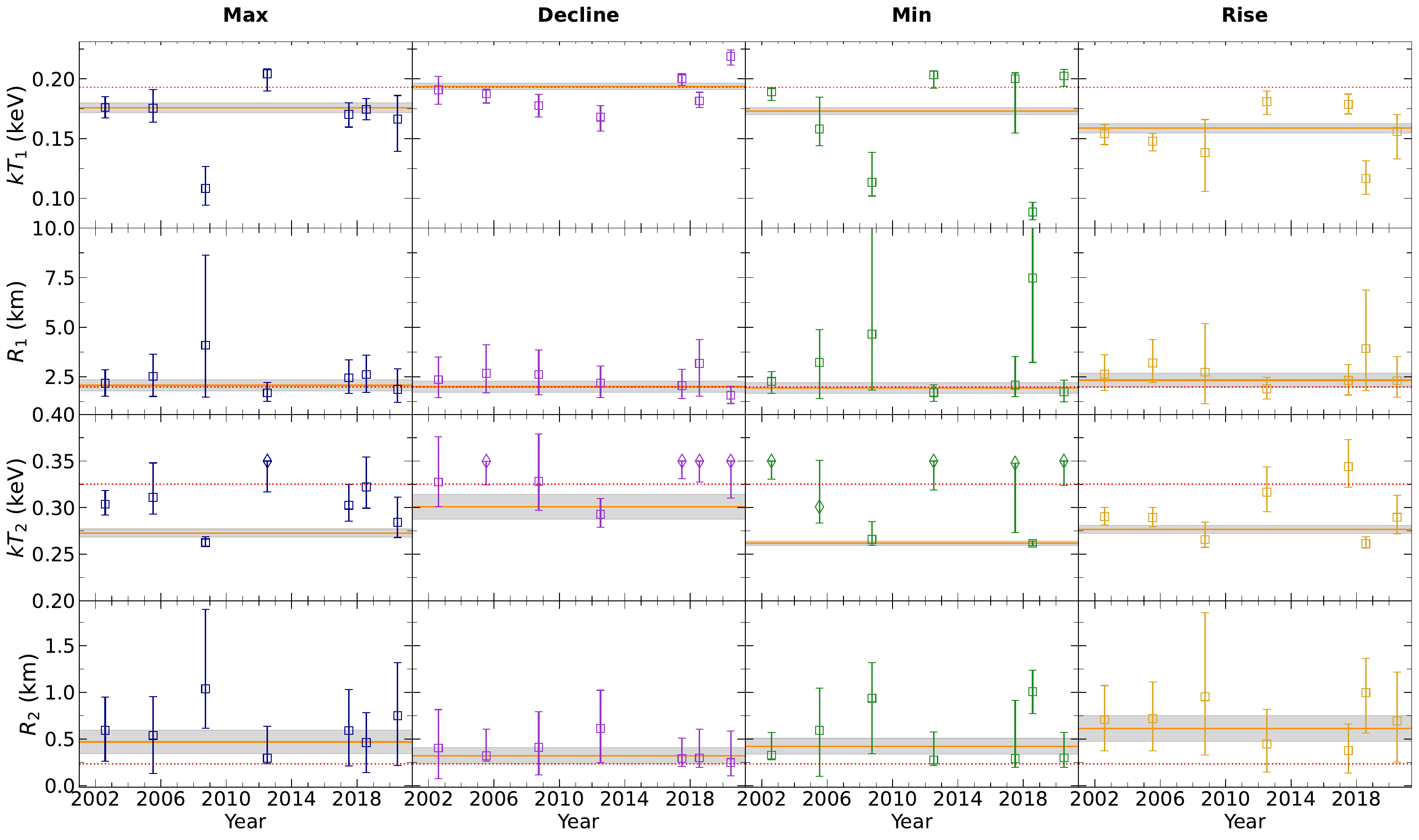}
\caption{
Continuum spectral model parameters as a function of time from fits to spectra at 7 epochs and 4 phase ranges (Max, Decline, Min, Rise; see Tables~\ref{tab:max}-\ref{tab:rise}).  Panels show blackbody emitting region temperatures $kT_1$ and $kT_2$ and radii $R_1$ and $R_2$. Errors are 1$\sigma$. Horizontal solid lines denote a fit of the data to a constant value, with shading indicating $1\sigma$ error of the fit, while  dotted lines are from a constant fit to the phase-averaged data (see Table~\ref{tab:constparamfit}).
\label{fig:contin}}
\end{figure*}

\begin{figure*}
\centering

\includegraphics[width=0.95\textwidth]{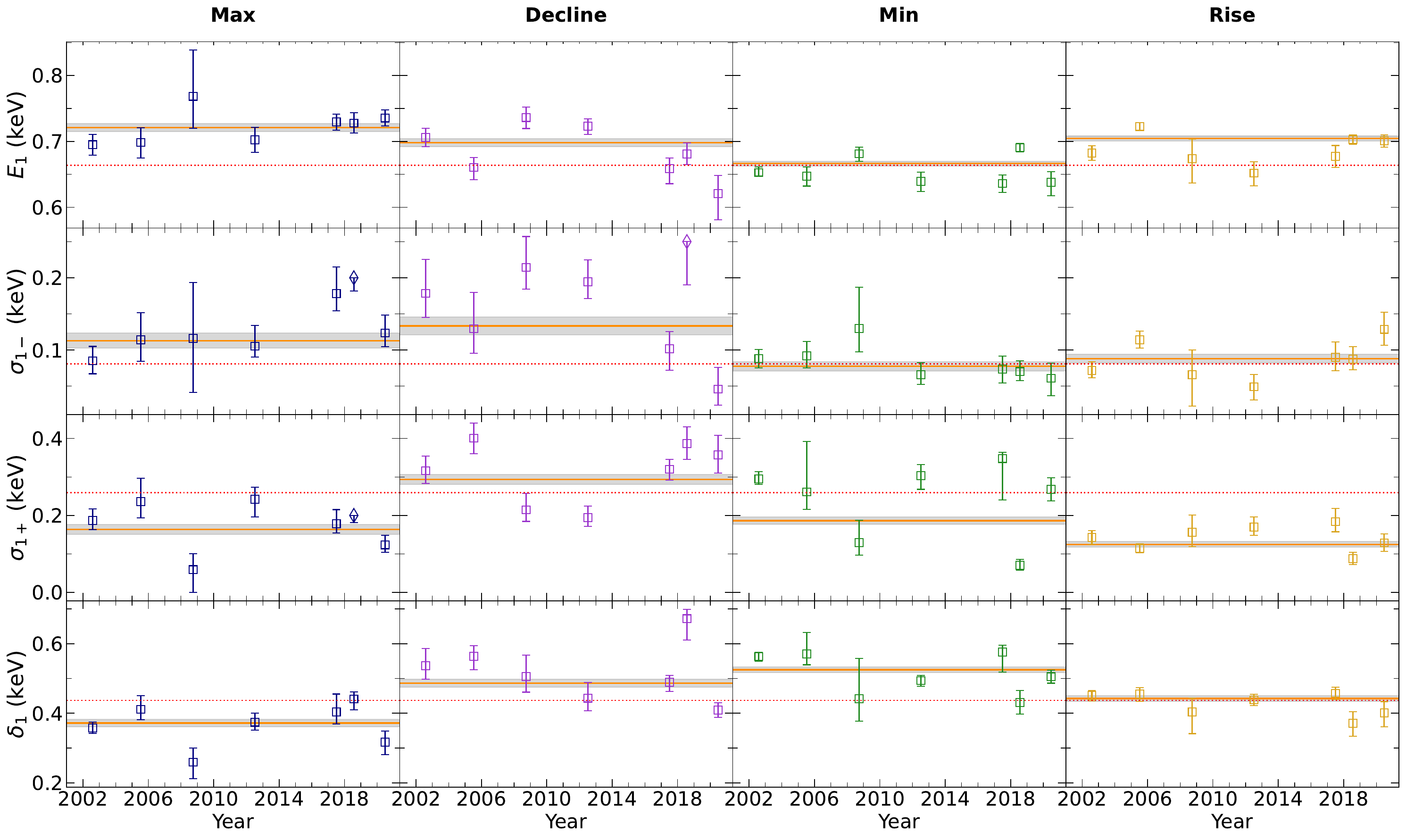}
\caption{Spectral model parameters for first absorption line as a function of time from fits to spectra at 7 epochs and 4 phase ranges (Max, Decline, Min, Rise; see Tables~\ref{tab:max}-\ref{tab:rise}).  Panels show line centroid energy $\enone$, widths below and above centroid $\sigmamone$ and $\sigmapone$, and depth $\linedone$.  Errors are 1$\sigma$. Horizontal solid lines denote a fit of the data to a constant value, with shading indicating $1\sigma$ error of the fit, while  dotted lines are from a constant fit to the phase-averaged data (see Table~\ref{tab:constparamfit}).
\label{fig:line1}}
\end{figure*}

\begin{figure*}
\centering
\includegraphics[width=0.95\textwidth]{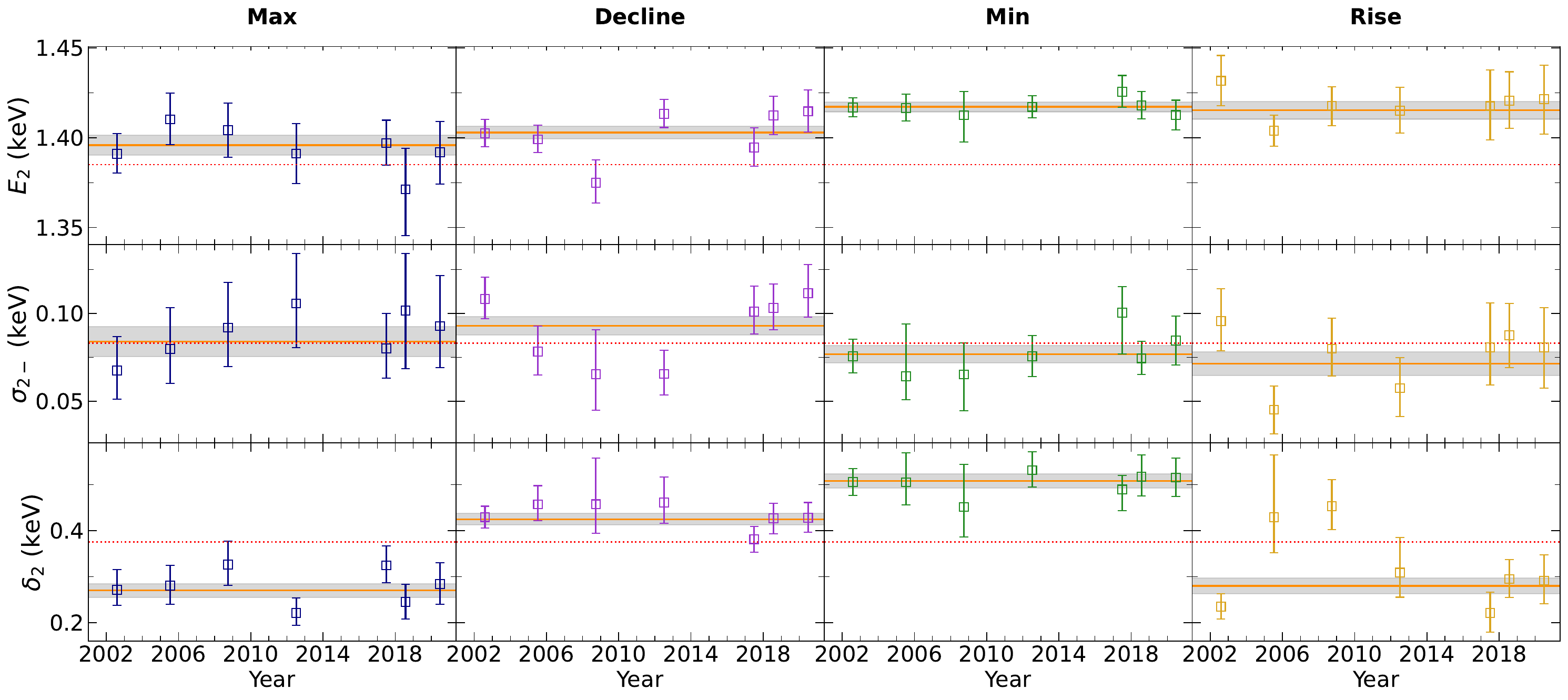}
\caption{
Spectral model parameters for second absorption line as a function of time from fits to spectra at 7 epochs and 4 phase ranges (Max, Decline, Min, Rise; see Tables~\ref{tab:max}-\ref{tab:rise}).  Panels show line centroid energy $\entwo$, half-width $\sigmamtwo$, and depth $\linedtwo$ (line is symmetric, such that $\sigmaptwo=\sigmamtwo$).  Errors are 1$\sigma$. Horizontal solid lines denote a fit of the data to a constant value, with shading indicating $1\sigma$ error of the fit, while  dotted lines are from a constant fit to the phase-averaged data (see Table~\ref{tab:constparamfit}).
\label{fig:line2}}
\end{figure*}

The resulting best-fits are satisfactory, with many having $\chi^2/\mbox{dof}\lesssim1.3$.
Some of the poorer fits are due to the lower data quality and apparent systematic features that our model fails to adequately capture.
For example, residuals at $\lesssim1\mbox{ keV}$ in the 2002, 2012, and 2017 spectra indicate the absorption line at $\sim0.7\mbox{ keV}$ is not simple in shape and potentially changes as a function of rotation phase, as also noted by \citet{DeLuca2004} in their analysis of the 2002 data,
and the possible Si instrumental edge at 1.839~keV mentioned above.
In addition, spectral model parameters are occasionally not well-constrained (e.g., low $kT_1$ and high $R_1$ in the 2008 maximum and minimum phases and 2018 minimum and rising phases),
and these can lead to structured residuals in the model fits.
These can be most easily seen in Figures~\ref{fig:contin}, \ref{fig:line1}, and \ref{fig:line2}, which plot the spectral parameters for the continuum, first absorption line, and second absorption line components, respectively, as functions of phase range and epoch.

Given the uncertainties and quality of fits,
none of the spectral parameters clearly demonstrate large variability over the 2002--2020 timespan of our phase-resolved data.
Therefore, we fit each parameter at a given phase range to a constant, with the results of these fits provided in the figures and in Table~\ref{tab:constparamfit}.
The continuum parameters in Figure~\ref{fig:contin} do not show significant changes with rotation phase, except for a possible shift in $kT_1$ of about 30~eV from declining to rising phases.
The most evident potential changes as a function of phase are in the parameters describing the first absorption line, as displayed in Figure~\ref{fig:line1}.
Here we see that the line energy decreases from 0.72~keV at maximum to 0.67 at minimum, accompanied by an increase in line depth from about 0.4~keV to 0.5~keV, as well as an increase in asymmetric width $\sigmapone$ from about 0.15~keV to 0.3~keV during the declining phase.
Meanwhile, the depth of the second absorption in Figure~\ref{fig:line2} seems to increase from 0.3~keV during the rising and maximum phases to 0.4~keV during the declining phase and then further to 0.5~keV during pulse phase minimum.

Finally, a prominent feature that stands out in a number of the declining phase spectra is what appears to be an absorption line at about 0.9~keV.
It is particularly evident in 2002, 2012, and 2017 in Figure~\ref{fig:each} and seems to be a distinct line separate from the 0.7~keV line.
The presence of a line at 0.9~keV leads to poor residuals as the fit tries to model both the 0.7 and 0.9~keV absorption lines as a single broad line at 0.7~keV.
A similarly larger line width at pulse decline was found by \citet{DeLuca2004} in their phase-resolved analysis of the 2002 pn spectra.
We discuss in Section~\ref{sec:mag} the implication on the magnetic field of \psr\ of this potential 0.9~keV absorption line appearing during certain rotation phases.

\section{Discussion} \label{sec:discuss}
\subsection{Magnetic field of \psr}
The primary objective of our study was to measure the two absorption lines
at approximately 0.7~keV and 1.4~keV in the X-ray spectrum of \psr\ and to determine
whether the energy of these lines changed during the two decades of \XMM\ EPIC pn data.
The source of these lines can be explained most naturally as electron
cyclotron harmonics, since the resulting magnetic field is approximately
the same as the field inferred from the pulsar's timing properties (see Section~\ref{sec:intro}).
The constancy or variability of these cyclotron absorption lines would
thus indicate whether the magnetic field of \psr\ is constant or variable.
While the timescales for magnetic field changes are generally much longer than a few decades (see, e.g., \citealt{Goldreich1992,Glampedakis2011}), phenomena observed from \psr\
suggest possible sudden changes in its magnetic field on short timescales, in particular, its possible
glitch in 2015 \citep{Gotthelf2018,Perez2026} and its recent radio detection despite earlier non-detections \citep{Zhang2026}.

We examined the total spectrum produced by combining the more than thirty
individual observations of \psr.
We also grouped in time these data from 2001 to 2020 to produce 9 epochs of rotation
phase-averaged spectra and 7 epochs of phase-resolved spectra.
We fitted these spectra with a model composed of two blackbodies and multiple (asymmetric in width) Gaussian absorption lines.
The results we obtained in some cases had relatively poor fit statistics, but based on the fit residuals and uncertainties, these are primarily due to the complex structure of the observed
absorption lines compared to the simple Gaussian shape of the assumed model.
Therefore, many of the conclusions we make below regarding the magnetic field
should be robust despite the fit quality.

\subsubsection{Surface magnetic field distribution and absorption line widths} \label{sec:mag}
Our measurements of the two strongest absorption lines in the \XMM\ pn spectra show that the best-fitting median energies of the lines $\enone$ and $\entwo$ do not change significantly from 2001 to 2020, but they might be changing as a function of rotation phase.
In more detail, $\enone$ ($\approx0.66\mbox{ keV}$) from the phase-averaged analysis declines by $\sim20\mbox{ eV}$ during the 19-year period (see top row in Figure~\ref{fig:avgline1}), but this small change is within uncertainties and $\entwo$ ($\approx1.38\mbox{ keV}$) and $\enone$ and $\entwo$ from the phase-resolved analysis do not show similar declines (see top row of Figures~\ref{fig:avgline2}, \ref{fig:line1}, and \ref{fig:line2}, respectively).
On the other hand, $\enone$ and $\entwo$ do change during a rotation cycle by $\sim50\mbox{ eV}$ and 20~eV, respectively, with the highest $\enone$ occurring at pulse maximum and the highest $\entwo$ shifted by half a cycle.
This variation could be caused by our view of the emitting region's magnetic field distribution changing as the star rotates, such that we see a change in the line energy $\Delta\enone/\enone\approx0.07$ or an effective magnetic field difference of $5\times10^9\mbox{ G}$.
Note that $\enone$ and $\entwo$ from the phase-resolved spectra are a few tens of eV higher than from the phase-averaged spectra; this is probably due to the lower-quality of the former spectra and our modelling of the complex-shaped lines with simple Gaussians.

We can obtain different insights
by examining the widths of the cyclotron absorption lines and their respective variations with epoch and rotation phase
since, as briefly sketched in Section~\ref{sec:model}, the widths of these spectral lines provide a measure of the range of magnetic field in the line-forming region.  Note however that we infer the magnetic field value from the cyclotron energy given by equation~(\ref{eq:cyc}) without accounting for the intrinsic width of the cyclotron resonance and conditions in the line-forming region and thus our derived magnetic field ranges are somewhat overestimated; a more accurate calculation is left for future work.
Now, like $\enone$ and $\entwo$, the total width $\sigmami+\sigmapi$ for both lines does not appear to change over the two decades of data (see middle panels of Figures~\ref{fig:avgline1}, \ref{fig:avgline2}, \ref{fig:line1}, and \ref{fig:line2}).
Our phase-averaged analysis yields $(\sigmamone+\sigmapone)/\enone\approx0.5$ or an emission spot with field strengths that span the range $(6.5-9.7)\times10^{10}\mbox{ G}$.

\begin{figure*}
\centering
\includegraphics[width=0.95\textwidth]{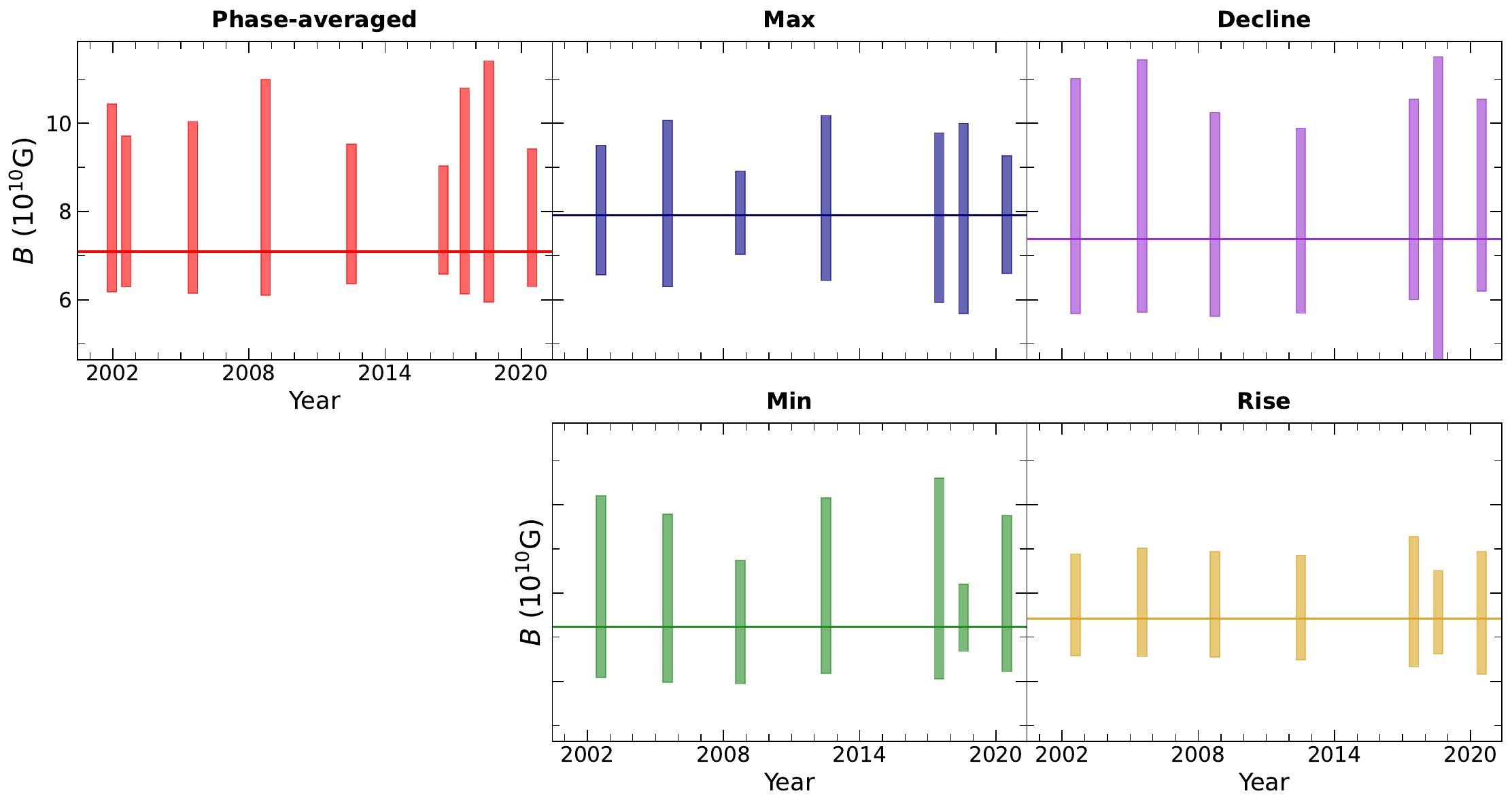}
\caption{Ranges of surface magnetic field $B$ contained within the hot spot on \psr\ as a function of time from fits to phase-averaged spectra at 9 epochs and phase-resolved spectra at 7 epochs and 4 phase ranges (Max, Decline, Min, Rise).  Each range is calculated assuming the absorption feature with full width from $\enone-\sigmamone$ to $\enone+\sigmapone$, where $\enone$ is centroid energy and $\sigmamone$ and $\sigmapone$ are widths below and above the centroid, is due to electron cyclotron resonance at energy $\Eb$.  Horizontal solid lines are calculated from the respective best-fit $\enone$ given in Table~\ref{tab:constparamfit}. 
\label{fig:mag}}
\end{figure*}

Meanwhile, the widths and depths of the lines clearly change with rotation phase, with the largest total widths occurring during the declining phase and the largest depths at pulse minimum.  This is the case for both lines but is most evident for $\sigmamone+\sigmapone$ and $\linedtwo$ (see middle rows of Figure~\ref{fig:line1} and bottom row of Figure~\ref{fig:line2}, respectively).
Of particular interest are the variations in line width because they provide the ranges of magnetic fields we can see as our view changes during a rotation cycle.
This is illustrated in Figure~\ref{fig:mag}, where we show the range in $B$ calculated from equation~(\ref{eq:cyc}) with $1+z_g=1.25$ and lower and upper bounds set by $E_B=\enone-\sigmamone$ and $\enone+\sigmapone$, respectively.
The largest width occurs during the declining phase and yields a range $B=(6.0-10.8)\times10^{10}\mbox{ G}$.

As discussed at the end of Section~\ref{sec:resultsphase}, the larger widths at pulse decline are due to the appearance in the spectra at this phase of a distinct absorption line at $\sim0.9\mbox{ keV}$ (see Figure~\ref{fig:each}, especially in 2002, 2012, and 2017).
Our spectral modelling fits this line together with the 0.7~keV line as a single absorption feature with a large total width ($\approx0.43\mbox{ keV}$).
In the context of the magnetic field of \psr, the appearance of the 0.9~keV line can be attributed to a higher magnetic field region of the NS surface (with $10^{11}\mbox{ G}$) moving into our field of view as the star rotates.
This changing view of the surface can be aided by relativistic light-bending and magnetic beaming effects (see, e.g., \citealt{Potekhin2014b}, for review).
We also note that the first harmonic of a cyclotron line at 0.9~keV would be at 1.8~keV,
which unfortunately overlaps with the Si instrumental edge at 1.839~keV and is unlikely to be distinguishable.
More thorough analysis of the 0.9~keV feature is necessary to determine whether this is an artifact of the analysis or a feature in the intrinsic spectra of \psr.

\subsubsection{Potential magnetic field change associated with 2015 glitch}
The phase-averaged and phase-resolved analyses presented here provide no evidence for a large change in the overall magnetic field of \psr\ during the 19~years of observations by \XMM. The only variability seen is that due to our changing view of the magnetic regions on the surface as the star rotates.  On the other hand, \citet{Perez2026} found, with $4\sigma$ significance, that the 2015 glitch produced a fractional change $\Delta\dot{f}/\dot{f}=-0.09$.
This would be expected to produce a 4.5~percent decrease in $B$, corresponding to a shift of $\sim30\mbox{ eV}$ and 60~eV in the cyclotron lines at the fundamental and first-harmonic, respectively. These shifts are on the order of the uncertainties in our measured line centroids $\sigma_{\enone}$ and $\sigma_{\entwo}$ but smaller than the measured line widths $\sigmami$ and $\sigmapi$, which allow variations in $B$ of several tens of percent, as described above. Consequently, spectral changes of the magnitude implied by the glitch cannot be distinguished from the uncertainties of the fitted line parameters. Therefore, while our analysis does not detect evidence for magnetic field evolution, it also cannot rule out changes of the (small) magnitude expected from the change in spin-down rate as a result of the glitch.

\subsection{Presence of absorption line near 2.1~keV}
In our phase-averaged and phase-resolved analyses, we found inclusion of a third absorption line at 2.1 or 2.2~keV led to improvements in the fit statistics for some epochs and rotation phases. Previous work by \citet{Mori2005} noted that, by changing the continuum model to include a magnetized hydrogen atmosphere or powerlaw component, a good fit to the 2002 data could be achieved without this third line. Additionally, their simulations and statistical analysis of a 2.1~keV feature did not yield significant detections in the pn, MOS1, and MOS2 data.  They surmised that spectral fit residuals in this energy range are most likely due to a Au~M~edge at 2.209~keV. 
A more detailed study like that conducted in \citet{Mori2005} with the much longer 2001--2020 data than just the 2002 data (623~ks  versus 152~ks) is beyond the scope of our paper but could provide more insights into the presence of the 2.1~keV feature and whether it is intrinsic to the spectrum of \psr.

\section{Conclusion} \label{sec:conclusion}
Continued monitoring of \psr\ is essential for understanding the long-term behavior of not only \psr\ but CCOs in general. With its recent detection in radio \citep{Zhang2026}, such monitoring of \psr\ could be done in radio, although its radio emission is weak and possibly transitory.  Thus future monitoring in X-ray would be useful too.  Measurements of its X-ray spectrum are also necessary.  The analysis presented here demonstrates that the X-ray spectral parameters of \psr\ remained stable over the period from 2001 to 2020, with no detectable variations in its surface magnetic field. This is in contrast to the glitch it probably experienced in its spin properties \citep{Gotthelf2018,Perez2026} and its recent emergence as a radio source, both of which could suggest a variable magnetic field that might manifest in X-ray spectra.

Several open issues remain for future investigations.
A more detailed statistical analysis of the 2.1~keV feature,
e.g., following the approach of \citet{Mori2005}, could better quantify its
significance.
Validation of the 0.9~keV (and 2.1~keV) absorption feature might be achieved
through analysis of \Chandra\ and \NICER\ data.
Maps of the surface temperature and magnetic field could follow from modelling
the asymmetric shapes of the absorption lines.
These detailed studies will be critical to shedding light on the mysterious properties that are present in CCOs.

\section*{Acknowledgements}
The authors thank C.-Y.~Ng for reading a draft and providing helpful comments.
The authors thank the anonymous referee for comments that led to improvements
in the paper.
TL acknowledges financial support from the Boughn-Gollub-Partridge Fund at Haverford College and the Haverford Koshland Integrated Natural Sciences Center (KINSC) Wintner/Love Fellowship.

This work is based on observations obtained with \XMM, an ESA
science mission with instruments and contributions directly funded
by ESA Member States and NASA.
This work uses data and software provided
by the High Energy Astrophysics Science Archive Research Center
(HEASARC), which is a service of the Astrophysics Science Division
at NASA/GSFC and High Energy Astrophysics Division of the
Smithsonian Astrophysical Observatory.
This work made extensive use of the NASA Astrophysics Data System
(ADS) Bibliographic Services and the arXiv.

\section*{Data Availability}

Data underlying this article will be shared on reasonable request
to the corresponding author.


\bibliographystyle{mnras}
\bibliography{1e1207}


\bsp
\label{lastpage}
\end{document}